\documentclass[aps,prd,eqsecnum,nofootinbib,superscriptaddress,floats,floatfix,tightenlines]{revtex4}
\usepackage[T1]{fontenc}
\usepackage{framed}

\usepackage{graphicx}
\usepackage{indentfirst}
\usepackage{latexsym}
\usepackage{multirow}
\usepackage{tabls}
\usepackage{color}
\usepackage{subfigure}

\usepackage{amsfonts}
\usepackage{amsmath}
\usepackage{bm}
\usepackage{mathrsfs}

\def\b{ \beta }

\def\b0{ {\bf 0} }

\def\lsim{\mathrel{\rlap{\lower3pt\hbox{\hskip1pt$\sim$}}
    \raise1pt\hbox{$<$}}}                
\def\gsim{\mathrel{\rlap{\lower3pt\hbox{\hskip1pt$\sim$}}
    \raise1pt\hbox{$>$}}}         

\def\coordeq{ \, \mathrel{ \rlap{\hbox{\hskip-2.5pt$=$} }
    \raise4pt\hbox{$\cdot$}} \, }                

\begin{document}

\title{A nonlinear scalar model of extreme mass ratio inspirals in effective field theory\\ I. Self force through third order}

\author{Chad R. Galley\footnote{chad.r.galley@jpl.nasa.gov}\footnote{Copyright 2010. All rights reserved.}}

\affiliation{Jet Propulsion Laboratory, California Institute of Technology, Pasadena, CA, 91109}
\affiliation{Center for Fundamental Physics, Department of Physics and\\ Center for Scientific Computation and Mathematical Modeling,\\University of Maryland, College Park, MD 20742}

\begin{abstract}
	The motion of a small compact object in a background spacetime is investigated in the context of a model nonlinear scalar field theory. This model is constructed to have a perturbative structure analogous to the General Relativistic description of extreme mass ratio inspirals (EMRIs). We apply the effective field theory approach to this model and calculate the finite part of the self force on the small compact object through third order in the ratio of the size of the compact object to the curvature scale of the background (e.g., black hole) spacetime. 
We use well-known renormalization methods and demonstrate the consistency of the formalism in rendering the self force finite at higher orders within a point particle prescription for the small compact object. This nonlinear scalar model should be useful for studying various aspects of higher-order self force effects in EMRIs but within a comparatively simpler context than the full gravitational case. These aspects include developing practical schemes for higher order self force numerical computations, quantifying the effects of transient resonances on EMRI waveforms  
and accurately modeling the small compact object's motion for precise determinations of the parameters of detected EMRI sources.
\end{abstract}

\maketitle

\section{Introduction}

The Laser Interferometer Space Antenna (LISA) \cite{LISA} is expected to see thousands of low-frequency gravitational wave sources when it is launched around $2025$. 
LISA's detection of extreme mass ratio inspirals (EMRIs) -- a white dwarf, neutron star or small black hole (collectively referred to as small compact object, SCO) inspiraling toward a supermassive black hole of mass $10^5 - 10^7$ solar masses -- is expected to provide an unprecedented level of insight into the structure of spinning black hole spacetimes as well as into the dynamics and populations of EMRIs in galactic nuclei.

EMRI signals will be extracted with matched filtering techniques from LISA's data stream. 
Once the presence of a signal is established one can use the resulting roughly-estimated parameters to refine the search using waveforms calculated with accuracies better than one cycle in the roughly $10^5$ cycles that are expected to accumulate during the last year of inspiral \cite{Gair:CQG21}. These high-accuracy templates are expected to be sufficiently accurate that the masses can be determined to about one part in $10^4$ \cite{BarackCutler:PRD69} and will require modeling the motion of the SCO with fractional accuracy of $10^{-5}$ or better. The SCO's motion and corresponding gravitational wave emission may be described using a perturbative treatment since the mass ratio for EMRIs is very small, between about $10^{-4}$ and $10^{-7}$ for EMRIs in LISA's detectable bandwidth.

Gravitational waves from EMRIs carry energy, linear momentum and angular momentum that results in a force on the compact object, called the self force, which causes the SCO to inspiral toward the supermassive black hole.
The self force on the SCO is conceptually different than, for example, the radiation reaction on a point charge in flat spacetime. Radiation reaction is local in time and is typically proportional to the time-derivative of the charge's acceleration. However, the self force also accounts for the history-dependent force arising from the interactions of the SCO with waves emitted in the past that have backscattered off the background spacetime curvature. As such, self force is nonlocal in time and depends on the past motion of the SCO. 
The self force can be computed perturbatively in powers of the mass ratio or, more generally, in powers of the size of the SCO to the curvature scale of the background spacetime it moves on.
To develop templates with sufficient accuracy to achieve LISA's science goals necessitates corrections through at least second order in the mass ratio. We motivate this last statement with a more detailed discussion below.

In this paper, we take first steps towards building high-accuracy EMRI waveforms by deriving the expressions for the self force on a SCO due to the emission of scalar perturbations. Specifically, we calculate the self force through third order in the mass ratio within a class of nonlinear scalar models that is constructed to be analogous to the kinematical structure of the perturbative General Relativistic description of EMRIs.

\subsection{High-accuracy EMRI waveforms}

Current theoretical techniques for generating high-accuracy waveforms are significantly underdeveloped when compared with those methods producing less accurate waveform templates, which are more useful for detecting EMRIs (for a review see \cite{Drasco:CQG23, Vallisneri:CQG26}). The former waveforms include ``Capra'' waveforms (see e.g., \cite{Drasco:CQG23}) and ``two-timescale'' waveforms \cite{Mino:PRD67, HindererFlanagan:PRD78}. Capra waveforms constitute the highest standard 
of accurate source modeling for EMRIs since the underlying calculations are based on a minimum number of assumptions and are thought to represent the binary's evolution and gravitational wave emission most accurately. These waveforms are sourced by the solutions to the self force equations of motion describing the perturbed motion of the SCO in the background supermassive black hole spacetime. The self force equation through first-order in the (very small) mass ratio is a complicated integro-differential equation, called the MiSaTaQuWa equation \cite{MinoSasakiTanaka:PRD55, QuinnWald:PRD56}, for the SCO's worldline coordinates. Accordingly, not a single Capra waveform has been computed despite recent advances and progress with numerical computations from several research groups; see Ref.\,\cite{Barack:CQG26} for a recent review. 

It was argued in \cite{Burko:PRD67, Rosenthal:PRD73} that if only the first-order (dissipative part of the) self force drives the quasi-circular inspiral of a SCO of mass $m$ moving in a Schwarzschild spacetime with mass $M$, then the accumulated phase of the gravitational waveform over an inspiral time $\sim M/\epsilon$, where $\epsilon = m/M \ll 1$, is, schematically,
\begin{align}
	\Phi \sim \frac{1}{\epsilon} + O(\epsilon^0) ~.
\label{phase1}
\end{align}
The last term represents the error in the phase from not including second and higher-order self force effects, which represents an $O(1)$ correction. Thus, to produce waveforms accurate to less than a cycle requires that second-order self force corrections be included in determining the SCO's motion.

Two-timescale waveforms are based on a systematic adiabatic expansion in which the typical orbital period $T_{\rm orb}$ is small compared to the inspiral timescale $T_{\rm insp}$. To leading order in $T_{\rm orb}/T_{\rm insp}$, Mino \cite{Mino:PRD67} showed that the waveform phase depends only on the time average of the dissipative part of the first-order self force. Hinderer and Flanagan \cite{HindererFlanagan:PRD78} extended Mino's work by placing it within a systematic two-timescale expansion to calculate ``post-adiabatic'' (PA) corrections to Mino's result. They found at 1PA that the time averaged dissipative part of the second-order self force is just as important as fluctuations in the conservative part of the first-order self force. Therefore, second-order self force corrections are important to maintain the consistency of the inspiral's adiabatic evolution.

\subsection{Transient resonances}

Recently, Flanagan and Hinderer \cite{Flanagan:2010cd} discovered that the SCO may undergo transient resonances during the course of its inspiral. Resonances occur only when the SCO evolves on a non-equatorial and eccentric orbit in a spinning supermassive black hole spacetime. Such orbital configurations are expected to be generic for EMRI sources detectable by LISA. During a transient resonance, the frequency of the true gravitational wave signal undergoes sudden jumps and an adiabatic treatment of the inspiral breaks down. Thus, the corresponding signal-to-noise ratio may be significantly diminished if using a template bank of two-timescale waveforms.

One way to address this problem is to provide increasingly accurate waveforms that are capable of tracking the phase evolution of the SCO even through the transient resonances.
There are at least two ways to do this.
One may patch a kludge waveform \cite{GairGlampedakis:PRD73, Babaketal:PRD75} (describing the resonant phases) into a very accurate two-timescale waveform (describing the system at all other times, during the adiabatic inspiral phases) or one may develop Capra waveforms that incorporate higher-order self force corrections to the SCO's motion.
The former approach is a reasonable possibility since the durations of the transient resonances are short compared to the radiation reaction time scale but will be only as accurate as the kludge model used. The latter method is the most direct and accurate way but may be difficult to realize given that even first order Capra inspiral waveforms have not yet been computed. Nevertheless, in the presence of resonances the accumulated phase of the waveform is, schematically,
\begin{align}
	\Phi \sim \frac{1}{\varepsilon} + \frac{ 1 }{ \sqrt{\varepsilon} } + O ( \varepsilon^0 )
\label{phase2}
\end{align}
and the $O(\varepsilon^{-1/2})$ contribution, which originates from passing through a resonance, requires knowing a part of the self force at second order in $\varepsilon$ \cite{Flanagan:2010cd}. Therefore, for EMRIs that pass through at least one resonance, the second and possibly higher-order contributions to the self force are especially important.

Estimates of the effects on the inspiral waveform phase from passing through a resonance indicate that the phase may change by $\sim 20$ rad for an EMRI with a mass ratio of $10^{-6}$ \cite{Flanagan:2010cd}. Not only is this a significant change for one resonant crossing but the effect accumulates for each resonance encountered during the inspiral. As a result, even detecting EMRIs with LISA could be affected by the dephasing from transient resonances. Thus, the motion of the SCO is needed with (possibly very) high accuracy in order to sufficiently describe its evolution before, during and after each transient resonance. 

Based on these previous works, there are several indications suggesting a need to model EMRI sources with high accuracy, which will require incorporating the effects on the SCO from second (and possibly higher) order self force corrections.

\subsection{Intermediate mass ratio inspirals}

Higher order self force corrections may also be needed to model binaries with less extreme mass ratios and could be useful even for those with comparable mass ratios. It is natural to think that including higher-order self force corrections in the SCO equations of motion will allow for the mass ratio to be relaxed to higher values. Doing so may offer the only way of describing binaries with intermediate mass ratios (IMRs), which have mass ratios in the range of $\sim 10^{-1} - 10^{-4}$, since neither the post-Newtonian approximation nor numerical relativity are particularly good tools for studying the inspirals of IMRs (however, see the recent work of \cite{Lousto:2010ut}). In fact, an alternative approach based on self force methods for binaries with IMRs may be useful 
for calibrating semi-analytical models (e.g., Effective One Body \cite{BuonannoDamour:PRD59}) and phenomenological hybrid waveforms \cite{Ajithetal:CQG24}.

\subsection{Self force in scalar models}

In this paper, we introduce a class of nonlinear scalar theories that is developed with a structure very similar to the perturbative General Relativistic description of EMRIs and will serve as a scalar analog of these sources. We calculate the finite (or regular) part of the scalar self force on the SCO through {\it third} order in the ratio of the size of the SCO -- $R_m$ -- to the background curvature length scale -- ${\cal R}$ -- and denoted by $\varepsilon = R_m / {\cal R}$, which is just the mass ratio in the strong-field regime of a supermassive black hole. We take the background to be specified for all time and, for simplicity, do not include the effects from the scalar field's stress-energy on the spacetime so that the scalar field is analogous to the propagation of metric perturbations on a fixed background spacetime. In regularizing the formally divergent self force expressions we use the standard tools of renormalization borrowed from the fields of high energy physics and condensed matter. The methods used in this paper will be of direct use for calculating the second order gravitational self force for EMRIs in a future paper.

Historically, scalar models offer a simpler framework for studying the underlying issues of self force regularization and for developing practical self force computational schemes. Indeed, the most useful regularization scheme (both in terms of physical intuition and practical computations) was first developed and understood in the context of the self force on a scalar charge from a linear scalar field in \cite{DetweilerWhiting:PRD67}. The first numerical computation of the self force (which was performed for a circular geodesic in Schwarzschild spacetime) was accomplished in a linear scalar theory in \cite{Burko:PRL84} and predated, by about seven years, the corresponding computation in the gravitational case \cite{BarackSago:PRD75}.

Because of the relative simplicities that scalar models afford, it seems likely that the physics of higher-order self force effects can be investigated more easily and quickly than in the gravitational EMRI context. Specifically, one can address the qualitative and quantitative effect that higher-order self force corrections have on the waveforms themselves (i.e., the change in the phase compared to first-order accurate waveforms, parameter estimation, etc.); how transient resonances and the number of resonances encountered during the inspiral affect the waveform and the SCO's motion; and how much the mass ratio can be relaxed to higher values while still having a reasonably accurate description of the system (the accuracy can be addressed using the next higher order to bound or estimate the errors). It is likely that these questions can begin to be studied using the results from this paper and with small modifications to some self force codes currently in use (particularly those using $3+1$ methods as in \cite{Vegaetal:PRD80} since these do not rely on a mode decomposition of the field).

\subsection{Previous work in higher-order self force corrections}

The first work (that we are aware of) regarding higher-order self force computations was carried out by Burko in Ref.\,\cite{Burko:PRD67}. He computed the second-order self force on a scalar charged particle in quasi-circular orbit in a Schwarzschild background. 
Burko derived a formal expression for the equations of motion on the particle for this scenario. The force on the particle included contributions from the second-order self force and also from the product of two first-order pieces. Since the second-order expressions were unknown, Burko retained only the latter contributions and estimated the change in the accumulated phase due to these self force corrections when compared to the first-order accurate waveform phase. He found that there was a relevant correction of $O(1)$ cycles to the first-order accurate phase. 
The change to the phase from amplitude corrections was found to be about a tenth of a cycle and hence irrelevant.

Later, Rosenthal developed a rigorous program to regularize scalar \cite{Rosenthal:CQG22} and gravitational \cite{Rosenthal:PRD72, Rosenthal:PRD73} perturbations through second order in perturbation theory. Using these regularized perturbations, he derived formal expressions for the second-order gravitational self force in Ref.\,\cite{Rosenthal:PRD74} in a gauge different from the standard Lorenz gauge, which he called the Fermi gauge \cite{Rosenthal:PRD72}. The Fermi gauge is a coordinate system in which
the regular part of the field perturbation and its covariant derivative are made to vanish on the worldline. As such, the first-order self-force corrections vanish in this gauge and the background spacetime is composed of the original background plus the first-order metric perturbations generated by the motion of the SCO. Unfortunately, Rosenthal's approach does not seem practical for numerical self force computations for EMRIs since constructing the Fermi gauge requires the first-order self force on the SCO to be made to vanish, which is accomplished by integrating the MiSaTaQuWa equation -- a feat that has yet to be accomplished.

Recently\footnote{Private communication with S.\,Detweiler.}, Detweiler \cite{Detweiler:CQGcapra} has shown with matched asymptotic expansions that the motion of the SCO can be described by a geodesic in a perturbed background spacetime having a smooth metric at the location of the particle through second order in $\varepsilon$. 
Detweiler's results imply that one may continue to interpret, at higher orders in $\varepsilon$, the SCO's motion as either being perturbed by self force corrections on a fixed background or as being geodesic in a perturbed spacetime.

\subsection{Organization}

This paper is organized as follows. In Section \ref{sec:eft} we give a brief overview of the effective field theory (EFT) approach as it applies to EMRIs. The EFT framework provides an efficient way to systematically calculate the self force at higher orders in $\varepsilon$. In Section \ref{sec:actionprinciple} we discuss how to consistently implement outgoing boundary conditions for the field within a variational principle. In Section \ref{sec:linearsf} we demonstrate how to isolate and evaluate the singular part of the well-known linear scalar self force on a charge in a non-vacuum spacetime so that we may compare the result from our formalism and methods to the standard result in Ref.\,\cite{Quinn:PRD62}. In Section \ref{sec:nonlinear} we develop a class of nonlinear scalar models that is designed to have a structure analogous to the perturbation theory used to describe EMRIs in General Relativity. In Section \ref{sec:nonlinearsf} we then compute the formally divergent self force expressions  in the nonlinear scalar model through third order in $\varepsilon$. In Section \ref{sec:renormalization} we renormalize these expressions by introducing counter terms into the action to cancel those divergences. In Section \ref{sec:eom} we write down the finite, third-order self force equations of motion. In Section \ref{sec:conclusion} we conclude with a discussion. The Appendices are devoted to deriving the quasilocal expansions used in regularizing the self force expressions in Section \ref{sec:nonlinearsf}, to proving that power-divergent integrals vanish in dimensional regularization and to listing the Feynman rules for the nonlinear scalar model introduced in Section \ref{sec:nonlinear}.

The regular part of the self force in the nonlinear scalar model valid through $O(\varepsilon^3)$ is given in (\ref{renormalizedsf1}) and (\ref{renormalizedsf3}) using the Detweiler-Whiting decomposition \cite{DetweilerWhiting:PRD67} for the retarded Green's function and in (\ref{renormalizedsf4}) and (\ref{effmass4}) using the Hadamard decomposition \cite{Hadamard}. Using the Detweiler-Whiting decomposition we find evidence suggesting that the self force through $O(\varepsilon^3)$ can be written solely in terms of the regular part of the field and its derivatives when evaluated on the worldline. In a later paper in this series we will explicitly show that this is indeed the case by calculating the radiative scalar perturbations and computing their effect on the SCO's motion \cite{Galley:Nonlinear2}.

We use units where $c=1$ and define the gravitational constant $G$ in terms of a mass parameter ($m_{pl}$) to be $32\pi G \equiv m_{pl}^{-2}$. The metric signature is $(-,+,+,+)$. We frequently use the notation where the worldline coordinates at proper times $\tau$ and $\tau'$ are denoted by the shorthand $z^\mu$ and $z^{\mu'}$, respectively, so that $z^\mu = z^\mu (\tau)$ and $z^{\mu'} = z^\mu (\tau')$. The same goes for tensors evaluated on the worldline at some proper time. We also use a mixed notation where a quantity such as $V(x; z]$ indicates that $V$ is a function of $x^\mu$ but is a functional of $z^\mu (\tau)$.

\section{Effective field theory approach}
\label{sec:eft}

In this paper we use the effective field theory (EFT) approach \cite{GoldbergerRothstein:PRD73} to calculate the self force equations of motion for a SCO interacting with a scalar field in a background curved spacetime. 
The SCO is described by a point particle on a worldline with coordinates $z^\mu(\tau)$.

As a description of gravitational EMRIs, one may worry about the applicability of the point particle approximation, particularly its validity for higher-order self force calculations, since it is known that point particle solutions do not exist in General Relativity \cite{GerochTraschen:PRD36}.
The EFT approach is consistent with this statement since the solutions for the metric perturbations generated by a point particle approximation of the SCO's motion are not applicable everywhere in the spacetime but are valid when the length scales being probed do not reach $R_m$ or smaller. In those cases, the EFT must be matched onto a more complete description for the internal structure of the SCO. However, for length scales larger than $R_m$ the point particle approximation is perfectly valid. In fact, the finite extent of the SCO can be parameterized in the EFT by including in the point particle action all terms that are consistent with the underlying symmetries: general coordinate invariance, reparametrization invariance of the worldline and local $SO(3)$ rotations (if the SCO is otherwise spherical when removed from all external influences). In this way, our ignorance of the SCO's internal structure is merely parameterized and the values of these new parameters are determined through matching calculations \cite{GoldbergerRothstein:PRD73}. In other words, the EFT description of the SCO dynamics essentially utilizes matched asymptotic expansions at the level of the action instead of the equations of motion, which is discussed further in Ref.\,\cite{KolSmolkin:PRD77}. 

Using EFT methods, it was shown in Ref.\,\cite{GalleyHu:PRD79} that moments induced on the SCO as it moves through the background spacetime exert a (tidal) force beginning at fourth order in $\varepsilon$ if the SCO is a black hole or neutron star. For a white dwarf, these finite size effects can be enhanced since a white dwarf is significantly larger than its gravitational radius. Hence, a white dwarf is more susceptible to tidal effects that, in turn, depend on how strongly curved the background spacetime is at the SCO's location. It was also shown in Ref.\,\cite{GalleyHu:PRD79} that a white dwarf undergoing some form of tidal disruption (i.e., Roche lobe overflow or tidal disintegration) may experience a tidal force that is numerically the same order of magnitude as a second-order self force correction. Furthermore, based on simple scaling arguments, the tidal disruption was found to occur outside the horizon of a supermassive black hole with mass $M$ when $M \gsim 1.4 \times 10^{-5} M_\odot$ (assuming the mass of the white dwarf is about $1.4 M_\odot$), which agrees within a factor of two with the standard lower bound of approximately $3 \times 10^{-5} M_\odot$ (see e.g., Refs\,\cite{Menouetal:NewAstron51, Sesanaetal:MNRAS391}). 

The EFT approach, as it applies to EMRIs, is introduced in Ref.\,\cite{GalleyHu:PRD79} and extends the original work of Goldberger and Rothstein \cite{GoldbergerRothstein:PRD73} who applied it to the post-Newtonian (PN) approximation for slow moving binary sources of gravitational waves. In the PN context, a significant amount of work has been done to calculate the conservative equations of motion for non-spinning \cite{GoldbergerRothstein:PRD73, GilmoreRoss:PRD78} and spinning compact objects \cite{Porto:PRD73, PortoRothstein:PRL97, PortoRothstein:0712.2032, PortoRothstein:PRD78, PortoRothstein:PRD78_2, Levi:PRD82, Perrodin:2010dy, Porto:CQG27, Levi:2010zu}. The formalism has been extended in Ref.\,\cite{GalleyTiglio:PRD79} to incorporate radiation reaction for generic orbits and to compute the waveforms that will be measured in gravitational wave interferometers; see also \cite{GoldbergerRoss:PRD81, Porto:2010zg} for the radiative multipole moments through 3PN for quasi-circular orbits with spinning compact objects. In addition, dissipative effects from the absorption of gravitational waves by the compact objects can also be treated in the EFT framework \cite{GoldbergerRothstein:PRD73_2, Porto:PRD77}. See also Refs.\,\cite{Cannellaetal:PRD80, Chu:PRD79} for applications beyond the standard two body problem in General Relativity.

Outside of the gravitational two-body context, the EFT approach has been used to derive the finite-size corrections to the radiation reaction on a charged object in Ref.\,\cite{GalleyLeibovichRothstein:PRL105}, to study cosmological perturbations in Ref.\,\cite{Baumann:2010tm}, to classify horizon geometries and topologies of higher dimensional black holes in Ref.\,\cite{Emparan_etal:PRL102}, and to study the Casimir interactions of objects in fluid surfaces and interfaces in Ref.\,\cite{YolcuRothsteinDeserno:1007.4760}, among other things.

The practical details of implementing the EFT approach in the context of EMRIs (Feynman diagrams, renormalization, etc.) will be given throughout the rest of this paper.

\section{A consistent action principle for open classical systems}
\label{sec:actionprinciple}

The EFT framework utilizes a perturbative treatment at the level of the action. Traditional approaches to the self force problem \cite{Quinn:PRD62, DeWittBrehme:AnnPhys9, MinoSasakiTanaka:PRD55, QuinnWald:PRD56} operate at the level of the equations of motion. We approach the self force calculation from an action principle for two important reasons. First, the subtraction of divergences is most easily handled by including counter terms in the action, which is particularly useful at higher orders in perturbation theory as we shall see in Section \ref{sec:renormalization}. 
 Second, an action principle unambiguously determines the appropriate worldline for the SCO. Simply put, it is the one that extremizes the action. In particular, at first order in $\varepsilon$ the worldline describing the motion of the compact object derived using the EFT approach \cite{GalleyHu:PRD79} is given by the self-consistent solution to the MiSaTaQuWa equation \cite{MinoSasakiTanaka:PRD55, QuinnWald:PRD56}. However, one may prefer to additionally expand the worldline in $\varepsilon$ so that the first-order self force corrections nudge the SCO away from the leading order geodesic motion. Such an approach is only valid until the so-called dephasing time when the corrections induce sufficiently large deviations from the original geodesic that the perturbation theory then breaks down. This has caused some confusion in the community about which worldline the self force corrects but it is clear using an action principle that the appropriate worldline is the one determined self-consistently with the self force, not the perturbed geodesic, which is also advocated in Ref.\,\cite{GrallaWald:CQG25}.

In this section, we show how to consistently implement retarded boundary conditions when integrating out the field from the action.
To provide a context, we derive the formally divergent self force in a linear scalar theory using the EFT approach. However, to make some of the details and issues more transparent we will not use Feynman diagrams in this section but instead calculate the self force step by step. We first attempt to derive the self force from the usual action principle, which ultimately fails to describe dissipation in the particle's dynamics, viz., the inspiral evolution. We then construct a new action principle that is capable of consistently and correctly incorporating the field's outgoing boundary conditions with the SCO's dynamics.
This formalism has been discussed previously in \cite{GalleyHu:PRD79, GalleyTiglio:PRD79} within the broader language of quantum field theory. Here, we motivate and develop this new action principle using only classical arguments.

\subsection{The usual action principle}

Consider a linear scalar field $\phi$ coupled linearly to a scalar charged particle with mass $m$ and charge $q$ in a non-vacuum background spacetime ($R_{\mu\nu} \ne 0$) so that the action for this system is given by
\begin{align}
	S[ z^\mu, \phi ] = - \frac{ 1}{ 2} \int_x ( \phi_{,\alpha} \phi^{,\alpha} + \xi R \phi^2 ) - m \int d\tau + q \int_x V(x;z] \phi 
\label{naiveaction1}
\end{align}
where $\int_x \equiv \int d^4x \, g^{1/2}$ is the integral with the invariant volume element, $\xi$ is a coupling constant and 
\begin{align}
	V(x; z] = \int d\tau \, \frac{ \delta^4 (x^\mu - z^\mu (\tau) ) }{ g^{1/2} } ~.
\label{Vee1}
\end{align}
To calculate the self force we first find the wave equation for the field by extremizing (\ref{naiveaction1}) with respect to $\phi$ only,
\begin{align}
	\Box \phi - \xi R \phi = - q V(x; z]  .
\end{align}
The solution with outgoing (retarded) boundary conditions is\footnote{We ignore homogeneous solutions throughout.}
\begin{align}
	\phi(x) = q \int_{x'} D_{\rm ret}(x,x')  V(x'; z] 
\label{retfield1}
\end{align}
where $D_{\rm ret} (x,x')$ is the retarded Green's function (or retarded propagator), which satisfies the following inhomogeneous wave equation,
\begin{align}
	\Box D_{\rm ret}(x,x') - \xi R (x) D_{\rm ret} (x,x') = - \frac{ \delta^4(x^\mu - x'^\mu) }{ g^{1/2}} ~.
\end{align}
Substituting the solution in (\ref{retfield1}) back into the action (\ref{naiveaction1}), which is what is meant by ``integrating out'' the field in this context, gives the {\it effective action} \cite{GoldbergerRothstein:PRD73, KolSmolkin:PRD77},
\begin{align}
	S_{\rm eff} [ z^\mu] = - m \int d\tau + \frac{ q^2 }{2} \int_x \int_{x'} V(x; z]  D_{\rm ret} (x,x') V(x'; z] ~.
\label{inouteffaction1}
\end{align}
Incidentally, it is this step (integrating out the field) that is performed indirectly by Feynman diagrams. Drawing the relevant Feynman diagrams at a given order in perturbation theory amounts to perturbatively solving the wave equation and substituting that solution back into the action. The advantage of using the diagrams is that one does not have to explicitly solve the wave equation order by order, which is particularly useful at higher orders in perturbation theory.

Calculating the equations of motion for the worldline (equivalently, the self force) from the effective action follows from the usual variational principle,
\begin{align}
	0 = \frac{ \delta S_{\rm eff} [z^\mu] }{ \delta z^\mu (\tau) }  
\end{align}
and yields, upon recalling the identity $D_{\rm ret} (x', x) = D_{\rm adv}(x,x')$,
\begin{align}
	m a^\mu = \frac{ q^2}{2} \int d\tau' \big( a^\mu + P^{\mu\nu} \nabla_\nu \big) \big( D_{\rm ret} (z^\mu, z^{\mu'}) + D_{\rm adv} (z^\mu, z^{\mu'}) \big)
\label{naivesf1}
\end{align}
where $P^{\mu\nu} = g^{\mu \nu} + u^\mu u^\nu$ and we have used the fact that
\begin{align}
	\frac{ \delta V (x; z] }{ \delta z^\mu (\tau) } = \big( a^\mu + P^{\mu\nu} \nabla_\nu \big) \frac{ \delta ^4 (x^\mu -z^\mu (\tau)) }{ g^{1/2} }  ~.
\label{dVeedz1}
\end{align}
Despite imposing retarded boundary conditions on the solution to the wave equation, the resulting effective action yields equations of motion for {\it conservative} particle dynamics since the Green's function appearing in (\ref{naivesf1}), $D_{ret} + D_{adv}$, is time-symmetric. This is simply a consequence of the fact that the full Lagrangian for the field-worldline system is time-reversal invariant and, as such, yields a variational principle for the worldline dynamics that is time-symmetric. 

If we couple the field to an auxiliary source $J(x)$ via a term in (\ref{naiveaction1}) of the form $\int_x J \phi$, then the field generated by the motion of the particle can be calculated from the effective action using
\begin{align}
	\phi ( x) = \frac{ \delta S_{\rm eff} [z^\mu , J] }{ \delta J(x) } \bigg|_{J=0}  
\end{align}
which here gives
\begin{align}
	\phi(x) = \frac{ q}{2} \int_{x'} \big( D_{\rm ret} (x, x') + D_{\rm adv} (x, x') \big) V ( x'; z] ~.
\label{retfield2}
\end{align}
Therefore, the effective action yields a radiative field that no longer respects the outgoing boundary conditions and disagrees with the explicit solution to the wave equation in (\ref{retfield1}). In fact, (\ref{retfield2}) describes radiation with no net flux of energy leaving the system, which is consistent with the absence of a dissipative component to the self force on the particle in (\ref{naivesf1}). 

It is worth remarking that for binaries with comparable masses in the post-Newtonian (PN) approximation the effective action is just the so-called Fokker action \cite{Fokker:ZPhys58, InfledPlebanski}. The Fokker Lagrangian has long been known to describe only conservative dynamics (see e.g., Ref.\,\cite{DamourSchafer:GRG17}). In Section \ref{sec:inin} we introduce a new action principle that essentially generalizes the Fokker action to allow for a description of dissipative processes as well as conservative ones. We have applied this new action principle in Ref.\,\cite{GalleyTiglio:PRD79} to derive the 2.5PN radiation reaction force of Burke and Thorne \cite{Burke:JMathPhys12, BurkeThorne:Relativity} from a Lagrangian formulation.

\subsection{A consistent action principle for open systems}
\label{sec:inin}

The shortcomings found in the usual action principle, as it applies to open systems, are caused by the time-reversal invariance of the full Lagrangian. 
One way to incorporate time-asymmetry into the true evolution of the system is to compare the actions of two different {\it histories} of the worldline and field variables, call them $(z^\mu_1, \phi_1)$ and $(z^\mu_2 , \phi_2)$. A useful comparison is to compute the difference in the actions of the two histories 
\begin{align}
	S [ z_1^\mu, z_2^\mu, \phi_1, \phi_2] \equiv S[ z_1^\mu, \phi_1] - S [ z_2^\mu, \phi_2 ] ~.
\label{ininaction1}
\end{align}
The interpretation of (\ref{ininaction1}) is as follows. Since the Lagrangian is time-reversal invariant it follows that the second term on the right side of (\ref{ininaction1}) is
\begin{align}
	- S [ z_2^\mu, \phi_2] = + \int_{t_f} ^{t_i} dt \, L [ z_2^\mu, \dot{z}^\mu_2, \phi_2, \partial_0 \phi_2 ]  ~.
\end{align}
Hence, the second history evolves {\it backward} through time from its final configuration to its initial one. This is why we have formed the difference of the actions -- (\ref{ininaction1}) naturally compares the forward time-evolution of the first history with the backward evolution of the second history. Thus, (\ref{ininaction1}) provides a measure of the time-asymmetry between the two histories. 

Obviously, if the two histories are the same, then (\ref{ininaction1}) vanishes identically -- if we compare the history evolved forward in time with the same history evolved backward in time then there is no difference. In addition, (\ref{ininaction1}) is a legitimate action in its own right since its extremal points correspond to extremal points of the actions for the two histories. 
However, there is only one relevant and physical history (for a given set of initial data), which we denote by $(z^\mu, \phi)$. Since the extremal points of the action determine the equations of motion then it is reasonable to require that the histories are equal to the physical one after performing the variations so that
\begin{align}
	\phi_1 & = \phi_2 = \phi  \label{phiend1} \\
	z^\mu_1 & = z_2 ^\mu = z^\mu ~.  \label{coordend1}
\end{align}
In this way, any time-asymmetries between the two histories will be captured in the description of the physical history's dynamics.

Incidentally, the action in (\ref{ininaction1}) is the one resulting from the classical limit of the corresponding quantum theory (of a quantum scalar field interacting with a point particle source) as derived using the so-called ``in-in'' formalism. The in-in formalism, first introduced by Schwinger \cite{Schwinger:JMathPhys2} and Keldysh \cite{Keldysh:JEPT20}, has been proven to give a causal description for open quantum systems. Hence, its classical limit provides a consistent implementation of the outgoing radiation boundary conditions. See Refs.\,\cite{GalleyHu:PRD79, GalleyTiglio:PRD79, CalzettaHu} and references therein for more details.

Let us now follow the same steps as in the previous section to calculate the self force but starting instead from the action in (\ref{ininaction1}). 
Written out fully (\ref{ininaction1}) is
\begin{align}
	S [z^\mu_{1,2}, \phi_{1,2} ] = {} & - m \int d\tau_1 + m \int d\tau_2  - \frac{1}{2} \int_x ( \phi_{1,\alpha} \phi_1 {}^{,\alpha} + \xi R \phi_1^2 ) + \frac{1}{2} \int_x ( \phi_{2,\alpha} \phi_2 {}^{,\alpha} + \xi R \phi_2^2 ) \nonumber \\
	& + q \int_x V(x; z_1] \phi_1 - q \int_x V( x; z_2]  \phi_2 
\label{ininaction2}
\end{align}
where $\tau_{1,2}$ are the proper times associated with the worldlines $z_{1,2}^\mu$.
It is convenient to make a change of variable to
\begin{align}
	\phi_- & = \phi_1 - \phi_2 \\
	\phi_+ & = \frac{1}{2} (\phi_1 + \phi_2)
\end{align}
in which case the action (\ref{ininaction2}) now reads
\begin{align}
	S[ z^\mu_{1,2}, \phi_\pm] = - m \int d\tau_1 + m \int d\tau_2  - \int _x ( \phi_{+, \mu} \phi_-{}^{,\mu} + \xi R \phi_+ \phi_- ) + q \int _x V_+ (x; z_{1,2}]  \phi_- + q \int_x V_- (x; z_{1,2}] \phi_+
\label{ininaction3}
\end{align}
where 
\begin{align}
	V_- (x; z_{1,2} ] & = V (x; z_1] - V( x; z_2 ] \\
	V_+ (x; z_{1,2} ] & = \frac{1}{2} \big( V(x; z_1 ] + V( x; z_2 ] \big) ~.
\end{align}
The wave equations for $\phi_\pm$ are
\begin{align}
	\Box \phi_\pm - \xi R \phi_\pm = - q V_\pm (x; z_{1,2}]  ~.
\end{align}
When reducing the histories to the physical one using (\ref{phiend1}) and (\ref{coordend1}) it follows that the field in the $\pm$ variables becomes $\phi_+ \to \phi$ while $\phi_- \to 0$. 
This implies that $\phi_+$ (i.e., the average of the two field histories) reduces to the physical field in this limit and the solution to the corresponding wave equation above must satisfy retarded boundary conditions. Therefore,
\begin{align}
	\phi_+ (x) = q \int_{x'} D_{\rm ret} (x,x') V_+(x'; z_{1,2}]  ~.
\label{retfield3}
\end{align}
Solving the $\phi_-$ wave equation for momentarily unspecified boundary conditions gives
\begin{align}
	\phi_- (x) = q \int _{x'} D (x,x') V_- (x' ; z_{1,2}] 
\label{advfield1}
\end{align}
for some Green's function $D(x,x')$ that will be determined shortly. Putting these solutions into the action (\ref{ininaction3}) gives the effective action,
\begin{align}
	S_{\rm eff} [ z^\mu _{1,2} ] = {} & -m \int d\tau_1 + m \int d\tau_2 + \frac{ q^2}{ 2} \int_x \int_{x'} V_- (x; z_{1,2} ] \big( D_{\rm ret} (x,x') + D(x', x) \big) V_+ (x' ; z_{1,2} ]   ~.
\label{inineffaction1}
\end{align}
The equations of motion for the particle follow by extremizing the effective action with respect to either $z_1^\mu$ or $z_2^\mu$ and setting $z_{1,2}^\mu = z^\mu$ using (\ref{coordend1}), which gives
\begin{align}
	m a^\mu = \frac{ q^2 }{ 2 } \int d\tau' \big( a^\mu + P^{\mu\nu} \nabla_\nu \big) \big( D_{\rm ret} (z^\mu, z^{\mu'}) +D (z^{\mu'}, z^{\mu}) \big)  ~.
\end{align}
In order for the equations of motion to exhibit the proper causal structure $D(x,x')$ must equal the {\it advanced} propagator,
\begin{align}
	D(x,x') = D_{\rm adv} (x,x')  
\end{align}
upon using $D_{\rm adv} (x', x) = D_{\rm ret} (x,x')$.
Therefore, the (formally divergent) self forced motion of the particle is given by
\begin{align}
	m a^\mu = q^2  \int d\tau' \big( a^\mu + P^{\mu\nu} \nabla_\nu \big) D_{\rm ret} (z^\mu, z^{\mu'})  ~.
\label{ppeom3}
\end{align}
We emphasize that what we have done here is to integrate out the field in the new action (\ref{ininaction1}) to obtain an effective action whose extremum describes particle motion {\it with} dissipative effects.

The fact that $\phi_-$ satisfies advanced boundary conditions (according to (\ref{advfield1})) is not a problem because the source of the $\phi_-$ field vanishes when setting $z_1^\mu = z_2^\mu = z^\mu$ implying that $\phi_-$ itself vanishes. Therefore, there is no contribution to any dynamics from advanced boundary conditions. 
For completeness, the effective action is
\begin{align}
	S_{\rm eff} [ z_{1,2}^\mu ] = - m \int d\tau_1 + m \int d\tau_2 + q^2  \int_x \int_{x'} V_- (x; z_{1,2} ]  D_{\rm ret} (x,x') V_+ (x'; z_{1,2} ]   ~.
\label{inineffaction2}
\end{align}
Note the similarities and differences between (\ref{inineffaction2}) and (\ref{inouteffaction1}).

Incidentally, coupling the fields $\phi_1$ and $\phi_2$ to external currents $J_1$ and $J_2$, respectively, computing either of $\delta S_{\rm eff} / \delta J_{1,2}$ and then setting $z_{1,2}^\mu =z^\mu$, $J_{1,2} =0$ gives the field radiated by the particle
\begin{align}
	\phi (x) = q \int _{x'} D_{\rm ret} (x,  x') V ( x' ; z]  ~,
\end{align}
which agrees with (\ref{retfield3}) when applying (\ref{phiend1}) and (\ref{coordend1}) and demonstrates the internal consistency of the formalism.

\subsection{Formal developments}
\label{sec:formaldevs}

Here, we briefly introduce some notation for the formalism developed in the previous section that will be used in the following sections to compute higher-order self force corrections in a nonlinear scalar model of EMRIs. 

The action in (\ref{ininaction2}) can be written in a condensed form by introducing a ``metric'' $c_{AB}$ used to raise and lower the $1,2$ indices labeling the histories,
\begin{align}
	c_{AB} = \left( \begin{array}{cc}
			1 & 0 \\
			0 & -1 
		\end{array} \right) = c^{AB} ,  
\end{align}
where a capital Roman letter takes values in $\{ 1,2 \}$. Then, (\ref{ininaction2}) can be written as
\begin{align}
	S [z_{1,2}^\mu ,  \phi_{1,2}] = - m \int d\tau_1 + m \int d\tau_2 - \frac{1}{2} \int _x ( \phi^A_{,\mu} \phi_A^{,\mu} + \xi R \phi^A \phi_A ) + q \int_x  V^A (x; z] \phi_A
\label{ininaction4}
\end{align}
with $V_A (x; z ] \equiv V (x; z_A]$, $\phi^1 = \phi_1$, $\phi^2 = - \phi_2$ and repeated indices indicate summation as usual so that $V^A = c^{AB} V_B$ and $\phi^A = c^{AB} \phi_B$.
Likewise, using the ``metric''
\begin{align}
	c_{ab} = \left( \begin{array}{cc}
			0 & 1\\
			1 & 0
		\end{array} \right) = c^{ab} 
\end{align}
where a lowercase Roman letter takes values in $\{ +,- \}$, allows for (\ref{ininaction3}) to be expressed more compactly as
\begin{align}
	S[ z^\mu_{1,2}, \phi_{1,2}  ] = - m \int d\tau_1 + m \int d\tau_2 - \frac{1}{2} \int_x ( \phi^a_{,\mu} \phi_a^{, \mu} + \xi R \phi^a \phi_a ) + q \int_x V^a (x; z]  \phi_a
\label{ininaction5}
\end{align}
with $V_- = V_1 - V_2$, $V_+ = (V_1 +V_2 ) / 2$, $\phi^+ = \phi_-$, and $\phi^- = \phi_+$.

Notice that (\ref{ininaction4}) and (\ref{ininaction5}) have the same form, suggesting that the action (\ref{ininaction1}) is, in a sense, ``covariant'' in the indices that label the variables. The transformation $\Lambda$ that goes from the $\pm$ basis to the $\{1,2\}$ basis is found by observing that
\begin{align}
	\left( \begin{array}{c}
		\phi_1 \\
		\phi_2 
	\end{array} \right) = \left( \begin{array}{c}
						\phi_+ +\frac{1}{2} \phi_- \\
						\phi_+ -\frac{1}{2} \phi_- 
					\end{array} \right)  = \left( \begin{array}{cc} 
										1 & \frac{1}{2} \\
										1 & - \frac{1}{2}
									\end{array} \right) \left( \begin{array}{c}
														\phi_+ \\
														\phi_-
													\end{array} \right)
\end{align}
or $\phi_A = \Lambda_A{}^a \phi_a$ where $\Lambda_A{}^a$ can be read off from the equation above. In addition, the metrics $c_{AB}$ and $c_{ab}$ are related by $\Lambda$ through
\begin{align}
	\phi_a \phi_b c^{ab} = 2 \phi_+ \phi_- =  \phi_1^2 - \phi_2^2 = \phi_A \phi_B c^{AB} =  ( \Lambda_A{}^a \phi_a )( \Lambda_B{}^b \phi_b) c^{AB}
\end{align}
from which it follows that
\begin{align}
	c^{ab} = c^{AB} \Lambda_A{}^a \Lambda_B{}^b
\end{align}
as could have been guessed from the ``covariant'' structure of the action with respect to the histories' labels.

The $\pm$ basis gives a convenient representation for the effective action in (\ref{inineffaction2}). To see this, define the following matrix of propagators
\begin{align}
	D^{ab} (x,x') = \left( \begin{array}{cc}
				0 & D_{\rm adv} (x,x') \\
				D_{\rm ret} (x,x') & 0
			\end{array} \right) ,
\label{mxofpropagators1}
\end{align}
which is symmetric under interchanges of indices and variables so that $D^{ab}(x,x') = D^{ba} (x', x)$ upon using the identity $D_{\rm adv} (x', x) = D_{\rm ret} (x,x')$. [Here, $D^{-+}(x,x') = D_{\rm ret} (x,x')$.] Then, (\ref{inineffaction1}) can be written in a more symmetrical form
\begin{align}
	S_{\rm eff} [ z_{1,2}^\mu ] = - m \int d\tau_1 + m \int d\tau_2 + \frac{q^2}{2} \int_x \int_{x'}  V_a (x;z]  D^{ab} (x,x') V_b(x'; z]    ~.
\label{inineffaction3}
\end{align}

\section{Self force regularization in a linear scalar theory}
\label{sec:linearsf}

The self force expression in (\ref{ppeom3}) is formally divergent when the retarded propagator is evaluated at $\tau'=\tau$. Hence, a suitable regularization procedure must be given to ensure a sensible and regular self force on the particle. In this section, we describe how we regularize the self force. We use the Hadamard decomposition of the retarded propagator into its so-called direct and tail pieces \cite{Hadamard} and then proceed to evaluate the singular parts of the proper time integral in (\ref{ppeom3}) using quasilocal expansions for $\tau'$ near $\tau$ (see Appendix \ref{app:expansions}). We demonstrate that our regularization procedure yields the correct finite self force expression as first derived by Quinn in \cite{Quinn:PRD62}. When calculating higher order self force corrections in Section \ref{sec:nonlinearsf} we use the Detweiler-Whiting decomposition \cite{DetweilerWhiting:PRD67} of the retarded Green's function as this has better behaved properties than Hadamard's form. The impatient reader who wants to get on with the nonlinear scalar model may safely skip ahead to Section \ref{sec:nonlinear}.

Whenever $x^\mu$ and $x^{\prime \mu}$ can be connected by a unique geodesic (the set of all such points connected to $x^\mu$ constitutes the normal neighborhood of $x^\mu$), the retarded propagator can be expressed as \cite{Hadamard, Poisson:LRR}
\begin{align}
	D_{\rm ret} (x,x') = \frac{1}{4\pi} \theta_+ (x, \Sigma_{x'}) \bigg( \Delta^{1/2}(x,x') \delta( \sigma(x,x')) + V(x,x') \theta( - \sigma(x,x') ) \bigg)  ,
\label{hadamard1}
\end{align}
which is the sum of a ``direct'' part (first term) and a ``tail'' part (second term).
Here, $\Sigma_{x'}$ represents a space-like hypersurface containing the point $x'$, $\theta(x, \Sigma_{x'})$ equals one if $x^\mu$ is to the future of $\Sigma_{x'}$ and zero otherwise, $\Delta(x,x')$ is the van Vleck determinant, $\sigma(x,x')$ is half the spacetime interval along the geodesic connecting the two points (also called Synge's world function) and $V(x,x')$ is a regular function. For notational convenience, let 
\begin{align}
	D_{\rm dir} (x, x' ) \equiv \frac{1}{4\pi} \theta_+ (x, \Sigma_{x'}) \Delta^{1/2} (x,x') \delta ( \sigma (x, x') ) 
\end{align}
be the direct part of the retarded propagator, which only has support on the future null cone of $x'^\mu$.

If $\tau_{\rm in}$ and $\tau_{\rm out}$ represent the proper times at which the worldline enters and exits, respectively, the normal neighborhood of the point $x^\mu = z^\mu (\tau)$ then we can split the integral in (\ref{ppeom3}) into contributions from outside and inside the normal neighborhood of $z^\mu (\tau)$,
\begin{align}
	F^\mu _{(1)}(\tau) & = q^2 \left( \int_{-\infty}^{\tau_{\rm in}} d\tau' + \int_{\tau_{\rm in}} ^{\tau_{\rm out}} d\tau' + \int_{\tau_{\rm out}}^\infty d\tau' \right) \big( a^\mu + P^{\mu\nu} \nabla_\nu \big) D_{\rm ret} (z^\mu, z^{\mu'}) ~. 
\label{firstorder2}
\end{align}
For the integration inside the normal neighborhood we may substitute $D_{\rm ret}$ from (\ref{hadamard1}) so that (\ref{firstorder2}) becomes
\begin{align}
	F^\mu _{(1)} (\tau) = {} & q^2 \int_{\tau_{\rm in}} ^{\tau_{\rm out}} {\hskip-0.1in} d\tau' \big( a^\mu + P^{\mu\nu} \nabla_\nu \big) \bigg[ D_{\rm dir} (z^\mu, z^{\mu'}) + \theta( \tau- \tau')  V(z^\mu, z^{\mu'})   \bigg] \nonumber \\
	& + q^2 \int_{-\infty}^{\tau_{\rm in}} \!\!\! d\tau' \big( a^\mu + P^{\mu\nu} \nabla_\nu \big) D_{\rm ret} (z^\mu, z^{\mu'} ) 
\label{firstorder3}
\end{align}
where we have used $\theta( - \sigma (z^\mu, z^{\mu'}) = 1$ for all $\tau'$. The contribution proportional to $V$ is regular and can be combined with the second line in (\ref{firstorder3}), which is also regular, so that
\begin{align}
	F^\mu _{(1)} (\tau) = {} & q^2 \int_{\tau_{\rm in}} ^{\tau_{\rm out}} {\hskip-0.1in} d\tau' \big( a^\mu + P^{\mu\nu} \nabla_\nu \big)  D_{\rm dir} (z^\mu, z^{\mu'})   + q^2 \lim_{\epsilon \to 0^+} \int_{-\infty}^{\tau - \epsilon} \!\!\! d\tau' \big( a^\mu + P^{\mu\nu} \nabla_\nu \big) D_{\rm ret} (z^\mu, z^{\mu'} ) 
\label{firstorder4}
\end{align}
where the limit is used to ensure that the second line receives no contribution from the singular coincidence limit of the propagator. All of the divergent structure is contained in the first integral of (\ref{firstorder4}), which we evaluate next.

There are two potentially divergent integrals that appear in (\ref{firstorder4}), which are proportional to
\begin{align}
	\int_{-\infty}^\infty d\tau' \, D_{\rm dir} (z^\mu, z^{\mu'}) {\rm ~~and~~} P^{\mu\nu} \int_{-\infty}^\infty d\tau' \, \nabla_\nu  D_{\rm dir} (z^\mu, z^{\mu'}) ~.
\label{twointegrals10}
\end{align}
Each integral has support only when $\tau'=\tau$ (which is why we have replaced the integration interval from $[\tau_{\rm in}, \tau_{\rm out}]$ to the whole real line) so that we can expand the integrands about $s \equiv \tau'-\tau = 0$. Using the expansions listed in Appendix \ref{app:expansions} it follows that
\begin{align}
	 \int_{-\infty}^\infty d\tau' \, D_{\rm dir} (z^\mu, z^{\mu'}) = \frac{1}{4\pi}  \int_{-\infty} ^0 ds \, \frac{ \delta (s) }{ | s | } =: \frac{ \Lambda}{4\pi},
\label{divintegral1}
\end{align}
which is completely divergent since there are no $O(s^0)$ terms in the product of $\Delta^{1/2} \delta (\sigma)$.

The second divergent integral in (\ref{twointegrals10}) can be calculated in a similar way. Using the relations in Appendix \ref{app:expansions} we find that
\begin{align}
	P^{\mu\nu} \int_{-\infty}^\infty d\tau' \, \nabla_\nu D_{\rm dir} (z^\mu, z^{\mu'}) = - \left( \frac{\Lambda}{4\pi} \right) \frac{ a^\mu }{2}  -  \bigg( \frac{1}{6} P^{\mu\nu} R_{\nu\alpha} u^\alpha + \frac{1}{3} P^{\mu\nu} \frac{ D a_\nu }{ d\tau} \bigg) \int_{-\infty} ^0 ds \, {\rm sgn}(s) \, \delta (s) ~.
\label{divintegral2}
\end{align}
The integrand in the second term is proportional to a product of distributions, ${\rm sgn}(s)$ and $\delta(s)$, suggesting that the integral is not well-defined. Specifically, the delta function has support only at $s=0$, which is where the ${\rm sgn}$ distribution is not defined \cite{Zemanian}. We will return to this issue shortly.

The regularized self force follows by substituting (\ref{divintegral1}) and (\ref{divintegral2}) into (\ref{firstorder4}) giving
\begin{align}
	F_{(1)}^\mu (\tau) = {} & \frac{q^2 a^\mu}{2}  \left( \frac{ \Lambda}{4\pi} \right) - \frac{ q^2}{4\pi} P^{\mu\nu} \bigg( \frac{1}{6} R_{\nu \alpha} (z^\mu) u^\alpha + \frac{ 1}{3}  \frac{ D a_\nu}{ d\tau} \bigg) \int_{-\infty}^0 ds \, {\rm sgn}(s) \, \delta (s) \nonumber \\
	& + q^2 \lim_{\epsilon \to 0^+} \int_{-\infty}^{\tau - \epsilon} \!\!\! d\tau' \big( a^\mu + P^{\mu\nu} \nabla_\nu \big) D_{\rm ret} (z^\mu, z^{\mu'} )  ~.
\label{firstorder5}
\end{align}
To evaluate the remaining $s$-integrals, we remark that the divergence arises when $s=0$ or, equivalently, $\tau' =\tau$. It is natural to parameterize the limit $\tau' \to \tau$ by shifting the argument of the delta function in (\ref{firstorder5}) from $s$ to $s + T$ for some $T \to 0^+$. This prescription also resolves the problem associated with (\ref{divintegral2}) since now the delta function has support at $s=-T$ where the ${\rm sgn}$ distribution is unambiguous. The two integrals in (\ref{firstorder5}) then evaluate to
\begin{align}
	\Lambda = \int_{-\infty}^0 ds \, \frac{ \delta (s ) }{ | s| } \longrightarrow \lim_{T \to 0^+} \int_{-\infty}^0 ds \, \frac{ \delta (s + T) }{ | s| } = \lim_{T \to 0^+} \frac{ 1}{T} \\
	 \int_{-\infty}^0 ds \, {\rm sgn}(s) \, \delta (s) \longrightarrow \lim_{T \to 0^+} \int_{-\infty}^0 ds \, {\rm sgn}(s) \, \delta (s+T) = -1
\end{align}
and the self force becomes
\begin{align}
	F_{(1)}^\mu (\tau) = {} & \frac{q^2 a^\mu}{8\pi}  \lim_{T \to 0^+} \frac{1}{T} + \frac{ q^2}{4\pi} P^{\mu\nu} \bigg( \frac{1}{6}  R_{\nu \alpha} (z^\mu) u^\alpha + \frac{ 1}{3}  \frac{ D a_\nu}{ d\tau} \bigg) + q^2 \lim_{\epsilon \to 0^+} \int_{-\infty}^{\tau - \epsilon} \!\!\! d\tau' \big( a^\mu + P^{\mu\nu} \nabla_\nu \big) D_{\rm ret} (z^\mu, z^{\mu'} )  ~.
\label{firstorder6}
\end{align}
The divergence in the self force is proportional to the acceleration. Hence, by introducing mass counter terms into the effective action (\ref{ininaction3}) of the form
\begin{align}
	- \delta _m \int d\tau _1 + \delta _m \int d\tau_2
\end{align}
we find that their contribution to the self force is $- \delta_m a^\mu$ and we can render (\ref{firstorder6}) finite if
\begin{align}
	\delta _m =  \frac{ q^2}{2} \frac{\Lambda }{ 4\pi} = \frac{q^2}{8\pi } \lim_{T \to 0^+} \frac{1}{T}  ~.
\end{align}

If instead of using the prescription $s \to s + T$ we used dimensional regularization, we would find that the divergent integral in (\ref{divintegral1}) actually {\it vanishes} so $\delta _m$ is zero. As is well known in the field of high-energy physics, an integral that diverges as a power of a cutoff (here, $T$ is the cutoff parameter) will vanish in dimensional regularization\footnote{The field that generates the divergence must be massless also.}. For the sake of completeness, we discuss in Appendix \ref{app:dimreg} how to use dimensional regularization in this context and verify our claim that (\ref{divintegral1}) vanishes.

Putting the pieces all together yields the (finite) self force equations of motion for the particle,
\begin{align}
	m a^\mu = \frac{ 1}{6}  \frac{ q^2}{ 4\pi} P^{\mu\nu} R_{\nu \alpha} u^\alpha + \frac{ 1}{3 }  \frac{ q^2}{ 4\pi} P^{\mu\nu} \frac{ D a_\nu}{ d\tau} + q^2 \lim_{\epsilon \to 0^+} \int_{-\infty} ^{\tau - \epsilon} \!\!\! d\tau' \, \big( a^\mu + P^{\mu\nu} \nabla_\nu \big) D_{\rm ret}  (z^\mu, z^{\mu'} )  ,
\label{eom1}
\end{align}
which is the expression first derived by Quinn in \cite{Quinn:PRD62}. From the equations of motion, one can define an effective mass for the particle as
\begin{align}
	m_{\rm eff} (\tau) = m - q^2 \lim_{\epsilon \to 0^+} \int_{-\infty}^{\tau-\epsilon} \!\!\! d\tau' \, D_{\rm ret} (z^\mu, z^{\mu'})
\label{effmass1}
\end{align}
from which the rate of change in the particle's mass is 
\begin{align}
	\frac{ d m_{\rm eff} }{ d \tau} = - \frac{1}{12} \frac{ q^2}{ 4\pi} (1 - 6 \xi ) R(z^\mu) - q^2 u^\alpha \lim_{\epsilon \to 0^+} \int_{-\infty}^{\tau - \epsilon} \!\!\! d\tau' \, \nabla_\alpha D_{\rm ret} (z^\mu, z^{\mu'} )  ~.
\label{mdot1}
\end{align}
See Ref.\,\cite{Poisson:LRR} for a general discussion and Ref.\,\cite{BurkoHartePoisson:PRD65} for an example of the scalar charge's mass evolving in a cosmological context. 

In the sections below, we will determine the higher order corrections to (\ref{eom1})--(\ref{mdot1}) but we must first generalize the action in (\ref{naiveaction1}) and (\ref{ininaction1}) to describe a nonlinear field theory model appropriate for extreme mass ratio inspirals.

\section{A nonlinear scalar model of EMRIs}
\label{sec:nonlinear}

There are many nonlinear scalar field theories that admit higher-order self force effects. However, we are interested in a class of models that possesses a structure analogous to the General Relativistic description of EMRIs. In particular, the motion of a SCO about a supermassive black hole is described by the perturbed Einstein-Hilbert Lagrangian (in the Lorenz gauge for the trace-reversed perturbations) about a black hole background spacetime (with metric $g_{\mu\nu}$ and dimensionless perturbations $h_{\mu\nu}/m_{pl}$) and is given schematically by\footnote{We have explicitly verified the structure of these higher order terms in the Einstein-Hilbert action through sixth order in $h_{\mu\nu}$, which is more than sufficient for our third order calculations here.}

\begin{align}
	S [ h_{\mu\nu} ] = - \sum_{n=2}^\infty \frac{1}{n! \, m_{pl}^{n-2} } \int_x  a_n^{\alpha \beta \mu_1 \cdots \mu_{2n}}(x) \nabla_\alpha h_{\mu_1 \mu_2} \nabla_\beta h_{\mu_3 \mu_4} h_{\mu_5 \mu_6} \cdots h_{\mu_{2n-1} \mu_{2n} }
\label{perturbedGR1}
\end{align}
where $m_{pl}^2 \equiv 1/( 32 \pi G)$ is a mass scale associated with the gravitational constant $G$ and the coefficients $a_n^{\alpha \beta \mu_1 \cdots \mu_{2n}}(x)$ are dimensionless and depend only on the vacuum background metric. 

One class of nonlinear scalar models that admits a structure analogous to (\ref{perturbedGR1}) is
\begin{align}
	S[ \phi] = - \sum_{n=2}^\infty \frac{1 }{n! \, m_{pl}^{n-2}} \int_x a_n^{\alpha \beta} (x) \partial_\alpha \phi \, \partial_\beta \phi \, \phi^{n-2} 
\label{nonlinearaction1}
\end{align}	
where $a_n^{\alpha \beta} = a_n g^{\alpha \beta}$ is proportional to the metric since that is the only dimensionless rank-2 tensor available to contract with the derivatives on the fields and we take $a_2 = 1$.

The action of a point particle coupled to gravitational perturbations in General Relativity is given by expanding the integral of the worldline's proper time on the perturbed spacetime,
\begin{align}
	S [ z^\mu, h_{\mu\nu} ] = - m \sum_{n=0}^\infty \frac{ 1}{ n! \, m_{pl}^n } \int d\tau \, b_n^{\mu_1 \cdots \mu_{2n}} (\tau) h_{\mu_1 \mu_2} (z^\mu) \cdots h_{\mu_{2n-1} \mu_{2n}} (z^\mu)
\label{perturbedGR2}
\end{align}
where the $b_n ^{\mu_1 \cdots \mu_{2n}} (\tau)$ depend only on the particle's four-velocity $u^\alpha(\tau)$ and $b_0 = 1$. In the EFT approach, extra terms describing the effects from the finite-size of the SCO can be included \cite{GoldbergerRothstein:PRD73}. However, we will leave the discussion of finite-size effects for a future paper. The analogous action for the interaction of the particle with the nonlinear scalar field is
\begin{align}
	S [ z^\mu, \phi ] = - m \sum_{n=0}^\infty \frac{1}{n! \, m_{pl}^n } \int d\tau \, b_n \phi^n (z^\mu)
\label{nonlinearpp1}
\end{align}
where the $b_n$ are constants and we take $b_0=1$. 

Combining (\ref{nonlinearaction1}) and (\ref{nonlinearpp1}) gives the full action for an analogous nonlinear scalar model of gravitational EMRIs
\begin{align}
	S [ z^\mu, \phi ] = - \frac{1}{2} \int_x \phi_{,\alpha} \phi^{,\alpha} A^2 (\phi / m_{pl} ) - m \int d\tau \, B(\phi / m_{pl} )
\label{nonlinear1}
\end{align}
where
\begin{align}
	A^2 (\phi / m_{pl} ) & = 1 + \sum_{n=1}^\infty \frac{ 2 a_{n+2}}{ (n+2)!} \left( \frac{ \phi }{ m_{pl} } \right)^n
\label{Asquared1} \\
	B ( \phi / m_{pl} ) & = 1 + \sum_{n=1}^\infty \frac{ b_n }{ n! } \left( \frac{ \phi }{ m_{pl} } \right)^n  ~.
\label{B1}
\end{align}
The wave equation for $\phi$ and the corresponding equations of motion for the particle are
\begin{align}
	\Box \phi & = - \frac{ A'}{A} \frac{\phi_{,\alpha} \phi^{,\alpha} }{ m_{pl} }  + \frac{m}{m_{pl}} \int d\tau \, \frac{ \delta^4 (x^\mu - z^\mu (\tau)) }{ g^{1/2}} \frac{ B' }{ A^2 } \\
	a^\mu & = - P^{\mu\nu} \nabla_\nu \ln B
\end{align}
where a prime denotes differentiation with respect to the function's argument, viz., $\phi / m_{pl}$.

\subsection{Field redefinition}

The action in (\ref{nonlinear1}) can be greatly simplified through the following field redefinition. Let $\psi (x)$ be given by
\begin{align}
	 \psi _{,\alpha}  =  \phi_{,\alpha}  A ( \phi / m_{pl} )
\label{psi1}
\end{align}
or $\psi = F( \phi / m_{pl} ) + C$ where $F(x) = \int dx \, A(x)$ is the anti-derivative of $A$ and $C$ is an integration constant that fixes $\psi=0$ when $\phi=0$. If $A$ cannot be integrated in closed form then 
one can integrate (\ref{psi1}) to find $\psi$ perturbatively in terms of $\phi$ using (\ref{Asquared1}) 
\begin{align}
	\frac{\psi}{ m_{pl}} = \frac{ \phi }{ m_{pl} } + \frac{a_3}{12} \left( \frac{ \phi }{ m_{pl} } \right)^2 + \frac{ 3 a_4 - a_3^2 }{ 216 } \left( \frac{ \phi }{ m_{pl} } \right)^3 + \cdots
\end{align}
and vice versa,
\begin{align}
 	\frac{ \phi }{ m_{pl} } = \frac{ \psi }{ m_{pl} } - \frac{a_3}{12} \left( \frac{ \psi }{ m_{pl} } \right)^2 - \frac{ 3 a_4 - 4 a_3^2 }{ 216} \left( \frac{ \psi }{ m_{pl} } \right)^3 + \cdots  ~.
\label{phiofpsi1}
\end{align}
The change of variables implies that the action in (\ref{nonlinearaction1}) becomes
\begin{align}
	S [ \psi ] = - \frac{1}{2} \int_x \psi_{,\alpha} \psi^{,\alpha}  ~.
\end{align}
The key point to note is that while we constructed (\ref{nonlinearaction1}) to mimic the nonlinear structure of perturbed General Relativity we have found that (\ref{nonlinearaction1}) can be transformed to a {\it linear} field theory, that is, one where the field does not interact with itself. This will have important implications regarding the divergent structure of higher-order self force effects in this model (see Section \ref{sec:renormalization}) and the corresponding renormalization of coupling constants.

A curious question naturally arises: If the nonlinear scalar theory is analogous to perturbed General Relativity then to what extent can one make a field redefinition to remove some, if not all, of the higher order ($n>2$) terms in (\ref{perturbedGR1})? We will address this in a forthcoming paper.

The full nonlinear scalar model for EMRIs is described by the following action upon using the change of variables from $\phi$ to $\psi$,
\begin{align}
	S [ z^\mu, \psi ] = - \frac{ 1}{2} \int_x \psi_{,\alpha} \psi^{,\alpha}  - m \int d\tau \, C (\psi / m_{pl} )
\label{nonlinear2}
\end{align}
where
\begin{align}
	C (\psi / m_{pl} ) & = 1 + \sum_{n=1}^\infty \frac{ c_n }{ n! } \left( \frac{ \psi }{ m_{pl} } \right)^n
\end{align}
and the first three $\{c_n\}$ coefficients in terms of the original coefficients, $\{a_n\}$ and $\{b_n\}$, are
\begin{align}
	c_1 &= b_1 \\
	c_2 & = b_2 - \frac{1}{6} a_3 b_1 \\
	c_3 &= b_3 - \frac{1}{2} a_3 b_2 - \frac{1}{12} a_4 b_1 + \frac{1}{9} a_3^2 b_1 ~.
\end{align}
The corresponding equations of motion for the field $\psi$ and the worldline are
\begin{align}
	\Box \psi & =  \frac{m}{m_{pl}} \int d\tau \, \frac{ \delta^4 (x^\mu - z^\mu (\tau) ) }{ g^{1/2} } C' (\psi / m_{pl} )
\label{waveeqn2} \\
	a^\mu & = - P^{\mu\nu} \nabla_\nu \ln C (\psi / m_{pl} )
\label{sfeom2}
\end{align}
where, as before, a prime denotes differentiation with respect to the argument of the function, viz., $\psi/ m_{pl}$.

\subsection{Power counting and Feynman diagrams}
\label{sec:pwrcounting}

The relevant scales in this nonlinear scalar model of EMRIs are the size of the SCO $R_m$, the curvature scale of the background spacetime ${\cal R}$ and the gravitational constant $m_{pl}$. These scales form a natural dimensionless parameter $\varepsilon \equiv R_m / {\cal R}$ that can be used to form a perturbative description for the EMRI whenever $R_m \ll {\cal R}$. If the SCO, with mass $m$, is in the strong-field region of a supermassive black hole, with mass $M$, then $\varepsilon$ is just the mass ratio $m/M$. Calculating the self force in the EFT approach using Feynman diagrams requires knowing which diagrams will appear at a given order in $\varepsilon$. This is accomplished by power counting, which determines the dimensional scaling of the  interactions in (\ref{nonlinear2}) in terms of the relevant scales in the theory.

The field $\psi(x)$ describes the radiation from the SCO implying that $\partial_\alpha \psi / \psi \sim 1/ {\cal R}$ and hence $x^\mu \sim {\cal R}$. Requiring that the kinetic term for the field is the leading order term in the perturbation theory (i.e., $\int_x (\partial \psi)^2 \sim O(\varepsilon^0)$) implies that $\psi \sim 1/ {\cal R}$.
The worldline integrals in (\ref{nonlinear2}) then scale as
\begin{align}
	- \frac{ m c_n }{ n! m_{pl}^n } \int d\tau \, \psi^n (z^\mu) \sim \frac{ m }{ m_{pl}^n } \frac{ 1}{ {\cal R}^{n-1}}  ~.
\label{powercounting1}
\end{align}
For $n=0$ this is $- m \int d\tau \sim m {\cal R} =: L$ and describes the leading order scaling for the SCO dynamics since the interactions with the field are perturbative corrections to this. In terms of $\varepsilon = R_m / {\cal R} \sim m / (m_{pl}^2 {\cal R} )$, (\ref{powercounting1}) becomes
\begin{align}
	- \frac{ m c_n }{ n! m_{pl}^n } \int d\tau \, \psi^n (z^\mu) \sim L \, \bigg( \frac{ \varepsilon}{ L } \bigg) ^{n/2}  ~.
\label{powercounting2}
\end{align}

In Appendix \ref{app:feynmanrules} we give the Feynman rules for the theory described by the action in (\ref{nonlinear2}). The vertices appearing in Feynman diagrams encode the interactions between the field and the SCO and are described by the worldline integrals in (\ref{nonlinear2}), which scale with $\varepsilon$ as shown in (\ref{powercounting2}). Therefore, to compute the effective action (and hence the self force) at O($\varepsilon^N)$ requires assembling diagrams with certain combinations of vertices such that the entire diagram scales as $\varepsilon^N$. These vertices are then connected by curly lines denoting the appropriate Green's function for the field (recall that there are two appearing in the matrix of propagators in (\ref{mxofpropagators1})). An example is shown in Figure \ref{fig:firstorder} for the Feynman diagram that contributes at first order in $\varepsilon$ to the effective action. Notice that, in this figure, the labels $a,b$ refer to the histories, discussed in Section \ref{sec:inin}.

\begin{figure}
\subfigure[Loop diagram]{
	\includegraphics[width=4.5cm]{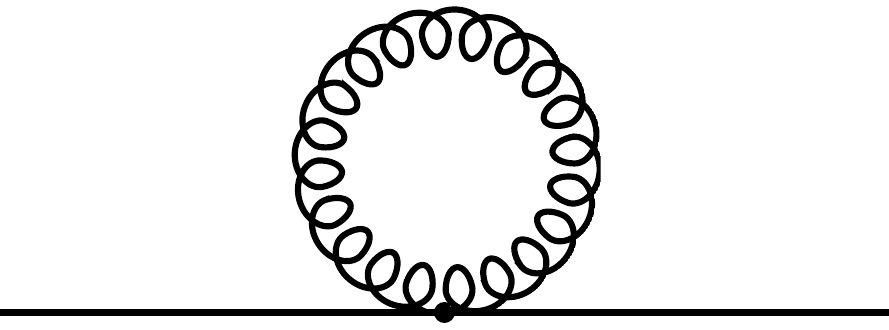} 
	\label{fig:loop}
}
\subfigure[Disconnected diagram]{
	\includegraphics[width=8cm]{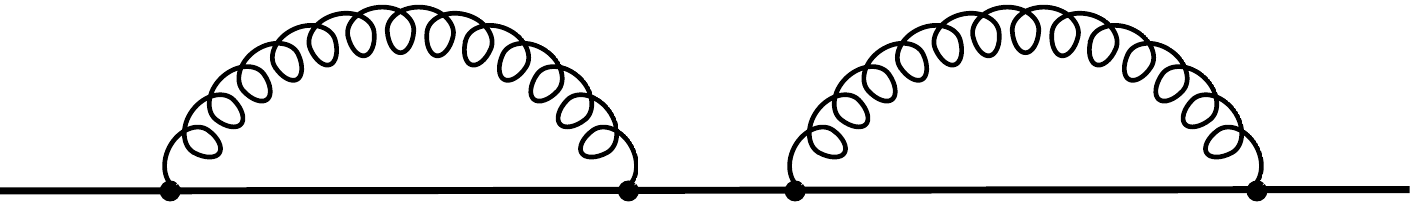}  
	\label{fig:disconnected}
}
	\caption{A loop diagram (a) and a disconnected diagram (b), neither of which can be generated by a local classical theory. The solid line represents the worldline of the small compact object and a curly line represents the field's propagator.}
\end{figure}

There are many possible diagrams one may draw. Since Feynman diagrams correspond to terms in the effective action, which are computed by perturbatively solving the wave equation and substituting that solution back into the full action, then the diagrams must be drawn according to that procedure. The solutions to the wave equation will never generate terms in the effective action involving $\sim \int d\tau \,D(z^\mu,z^\mu)$, for example. Such terms would appear as loops of curly lines (see Figure \ref{fig:loop}) and are not admissible in a classical theory such as (\ref{nonlinear2}). In fact, diagrams with such loops of curly lines turn out to be quantum effects and are suppressed in the effective action by powers of $\hbar/L$, as discussed in \cite{GalleyHu:PRD79}. Therefore, all diagrams here must not contain any loops of curly lines.

In addition, all diagrams must be {\it connected} in the sense that cutting the worldline at any place leaves the resulting diagram connected by curly lines. Classical solutions to the wave equation cannot generate terms in the effective action corresponding to disconnected diagrams. To see this, let's assume they did. Then there could be terms appearing in the effective action like the one in Figure \ref{fig:disconnected}, and schematically written out here as
\begin{align}
	 \sim \bigg[ \int d\tau \int d\tau' \, D (z^\mu, z^{\mu'}) \bigg]^2  ~,
\end{align}
that could only come from the following term in the full action
\begin{align}
	\bigg[ \int d\tau \, \psi (z^\mu) \bigg]^2 = \int d\tau \int d\tau' \, \psi(z^\mu) \psi(z^{\mu'}) ~.
\end{align}
Clearly, such terms are not present in (\ref{nonlinear2}) since the Lagrangian is {\it local} in time. Therefore, such terms also cannot appear in the effective action and the relevant Feynman diagrams must be connected ones.

These two conditions together (no loops or disconnected diagrams) imply that every diagram contributing to the effective action scales as a single power of $L = m {\cal R}$. Essentially, these rules come about because (\ref{nonlinear2}) describes a local, classical field theory.

\section{Self force in the nonlinear scalar model}
\label{sec:nonlinearsf}

In this section we calculate the expressions for the self force on the compact object through {\it third} order in $\varepsilon$ within the nonlinear scalar model of (\ref{nonlinear2}), i.e., after implementing the field redefinition. The Feynman diagrams that arise at each order in this perturbation theory have the same topology as a subset of those diagrams appearing in the gravitational case describing the dynamics of physical EMRIs. As a result, the computational tools and methods used and developed here will be indispensable for calculating the gravitational self force at higher orders in the Lorenz gauge in a future paper.

To show that our renormalization procedure works for singularities arising from nonlinear interactions of point sources we will explicitly retain all divergences until we are finished computing the third-order self force. At that point we will shall identify the counter terms that cancel these divergences and render the self force expressions finite.

\subsection{Self force at first order}
\label{sec:firstorder}

\begin{figure}
	\includegraphics[width=4cm]{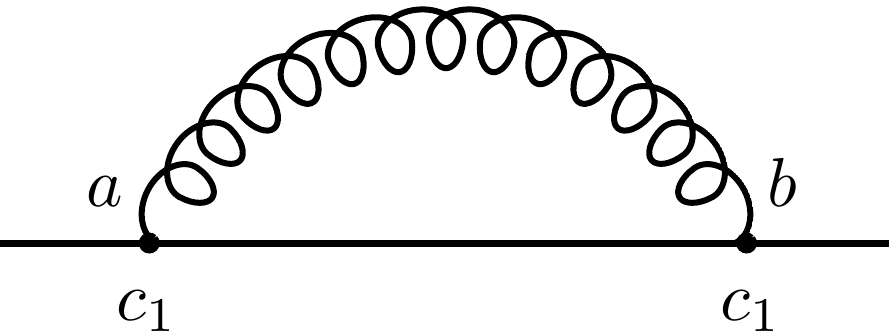}
	\caption{The Feynman diagram contributing to the effective action at first order in $\varepsilon$. Power counting indicates that this diagram scales as $\sim \varepsilon L$.}
	\label{fig:firstorder}
\end{figure}

At first order there is a single Feynman diagram contributing to the effective action, which is shown in Figure \ref{fig:firstorder}. We have already regularized the first-order self force in Section \ref{sec:linearsf}. In that section we decomposed the retarded propagator into its direct and tail pieces in order to obtain a sufficiently regular self force expression to compare with the known result in \cite{Quinn:PRD62}. However, that decomposition is not particularly useful for practical self force computations, especially at higher orders, as we shall see in Section \ref{sec:eom}.

A better decomposition is provided by the Detweiler-Whiting prescription \cite{DetweilerWhiting:PRD67} where the retarded propagator is split into a regular ($R$) and singular ($S$) part,
\begin{align}
	D_{\rm ret} (x,x') = D_R (x,x') + D_S (x,x')  ~.
\end{align}
The advantage of this decomposition is that the regular part $D_R$ satisfies the {\it homogeneous} wave equation, is regular on the worldline and is the part of $D_{\rm ret}$ actually responsible for forcing the SCO.  Therefore, the corresponding contribution to the self force is regular on the worldline and one can make sensible predictions. The singular part $D_S$ satisfies the {\it inhomogeneous} wave equation, carries all the divergent structure of $D_{\rm ret}$ and exerts no force on the SCO.
When $x$ and $x'$ are in a normal neighborhood we can write the regular and singular parts as
\begin{align}
	D_R (x,x') = {} & \frac{1}{8\pi} \Delta^{1/2} (x,x') \delta \big( \sigma  \big) \big[ \theta_+ (x, \Sigma_{x'}) - \theta_- (x, \Sigma_{x'} ) \big] + \frac{1}{4\pi} V(x,x') \left[ \theta_+ (x, \Sigma_{x'}) \theta (- \sigma) + \frac{1}{2} \theta (\sigma) \right]
\label{regular1} \\
	D_S (x,x') = {} & \frac{1}{8\pi} \Delta ^{1/2} (x,x') \delta (\sigma) - \frac{1}{8\pi} V( x,x') \theta (\sigma) ~.
\label{singular1}
\end{align}
See \cite{Poisson:LRR} for a pedagogical presentation and discussion of the Detweiler-Whiting decomposition.

Following steps similar to those implemented in Section \ref{sec:linearsf} (but using the Detweiler-Whiting decomposition instead), we isolate the singular and regular parts of the integrals in (\ref{ppeom3}) so that 
\begin{align}
	\int d\tau' \, D_{\rm ret} (z^\mu, z^{\mu'}) & = \frac{\Lambda}{4\pi} + I_R (z^\mu) 
\label{divintegral3} \\
	P^{\mu\nu} \int d\tau' \, \nabla_\nu D_{\rm ret} (z^\mu, z^{\mu'}) & = - \frac{\Lambda}{4\pi} \frac{a^\mu}{2} + P^{\mu\nu} \nabla_\nu I_R (z^\mu)
\label{divintegral4}
\end{align}
where $I_R$ is defined as
\begin{align}
	I_R (x^\mu) & \equiv  \int_{-\infty}^\infty d\tau' \,  D_{R} (x^\mu , z^{\mu'}) 
\end{align}
and, by construction, is regular on the worldline so that $\nabla_\nu I_R(x^\mu)$ is well-defined when $x^\mu \to z^\mu$ \cite{DetweilerWhiting:PRD67}. The regularized first-order self force is then 
\begin{align}
	F^\mu _{(1)} (\tau) = {} & \frac{ m^2 c_1^2}{ 2 m_{pl}^2 } \frac{\Lambda}{4\pi} a^\mu + \frac{ m^2 c_1^2}{ m_{pl}^2} \big( a^\mu + P^{\mu\nu} \nabla_\nu \big) I_R (z^\mu)  ~.
\label{firstorder7}
\end{align}
Notice that (\ref{firstorder7}) agrees with our earlier derivation in Section \ref{sec:linearsf} based on Hadamard's decomposition of the retarded Green's function, which is expressed in terms of direct and tail pieces. To see this, we identify $q = m c_1 /m_{pl}$ so that the tail part of the self force in (\ref{firstorder6}) is related to the Detweiler-Whiting regular part by
\begin{align}
	(a^\mu + P^{\mu\nu} \nabla_\nu) I_R(z^\mu) = \frac{1}{4\pi} P^{\mu\nu} \bigg( \frac{ 1}{6} R_{\nu\alpha} u^\alpha + \frac{ 1}{3} \frac{ D a_\nu }{ d\tau} \bigg) + \lim_{\epsilon \to 0^+} \int_{-\infty}^{\tau-\epsilon} \!\!\! d\tau' \, (a^\mu + P^{\mu\nu} \nabla_\nu ) D_{\rm ret} (z^\mu, z^{\mu'})
\end{align}

\subsection{Self force at second order}

\begin{figure}
	\includegraphics[width=7cm]{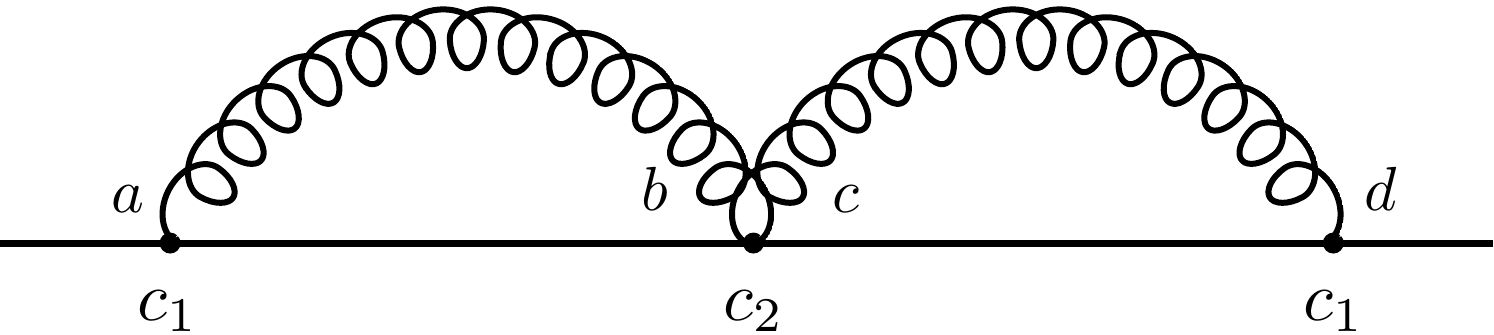}
	\caption{The diagram contributing to the effective action at second order in $\varepsilon$. Power counting indicates that this diagram scales as $\sim \varepsilon^2 L$.}
	\label{fig:secondorder}
\end{figure}

At second order in $\varepsilon$, power counting indicates that one diagram contributes to the effective action, which is shown in Figure \ref{fig:secondorder} and scales as $\varepsilon^2 L$. 
The Feynman rules for that diagram imply that the corresponding terms in the effective action are
\begin{align}
	S^{(2)}_{\rm eff} [ z_{1,2}^\mu] = \frac{1}{2!} \int_x \int_{x'} \int_{x''} T_a (x; z] D^{ab} (x,x') T_{bc} (x'; z] D^{cd} (x', x'') T_{d} (x''; z]  ~.
\label{secondeffaction1}
\end{align}
Here, we use the notation from Appendix \ref{app:feynmanrules} so that a vertex with $n$ scalar fields attached to the worldline is represented by
\begin{align}
	T_{a_1 \cdots a_n} (x; z] = - \frac{ m c_n }{ m_{pl}^n} c_{a_1 \cdots a_n} {}^b V_{b } (x; z]
\label{vertexT1}
\end{align}
where $c_{a_1 \cdots a_n}{}^b = \Lambda_{a_1} {}^{A_1} \cdots \Lambda_{a_{n+1}}{}^{A_n} \Lambda^b {}_B c_{A_1 \cdots A_{n}}{}^B$ and 
\begin{align}
	c_{A_1 \cdots A_p} = \left\{ \begin{array}{cl}
							1 & A_1 = \cdots = A_p = 1 \\
							-1 & A_1 = \cdots = A_p = 2 \\
							0 & {\rm otherwise}
						\end{array} \right.  ~.
\end{align}	
Substituting in (\ref{vertexT1}) into (\ref{secondeffaction1}) gives
\begin{align}
	S^{(2)}_{\rm eff} [ z_{1,2}^\mu ] = \frac{1}{2} \bigg( - \frac{ m c_1}{ m_{pl}} \bigg)^2 \bigg( - \frac{ m c_2 }{ m_{pl}^2 } \bigg) \int_x \int_{x'} \int_{x''} V_a (x; z] D^{ab} (x,x') c_{bc}{}^e V_e (x'; z] D^{cd} (x', x'') V_d (x''; z]
\end{align}
We perform the contractions in the $\pm$ basis and use the non-zero elements of the $c$-tensors,
\begin{align}
	c_{+-}{}^+ & = c_{+--} = 1 \\	
	c_{-+}{}^+ & = c_{-+-} = 1 \\
	c_{--}{}^- & = c_{--+} = 1 \\
	c_{++}{}^- & = c_{+++} = 1/4 ~,
\end{align}
to find that the effective action becomes
\begin{align}
	S ^{(2)}_{\rm eff} [z_{1,2}^\mu ] = {} & -  \frac{ m^3 c_1^2 c_2 }{ 2 m_{pl} ^4}  \int_x \int_{x'} \int_{x''} \bigg\{ 2 V_- (x; z] D_{\rm ret} (x,x') V_+(x' ; z] D_{\rm ret} (x', x'') V_+( x''; z] \nonumber \\
	& {\hskip1in} + V_-(x; z] D_{\rm ret}(x, x') V_+ (x'; z] D_{\rm ret} (x, x'') V_+ (x''; z] + O( V_-^2 ) \bigg\}  ~.
\label{tempS2eff1}
\end{align}
The self force at second order in $\varepsilon$ is then found by varying (\ref{tempS2eff1}) with respect to either $z_1^\mu$ or $z_2^\mu$ using the identity in (\ref{dVeedz1}) and then setting $z_{1,2}^\mu = z^\mu$ according to (\ref{coordend1}),
\begin{align}
	F_{(2)}^\mu (\tau) = {} & - \frac{ m^3 c_1^2 c_2 }{ 2 m_{pl}^4 }  \big( a^\mu + P^{\mu\nu} \nabla_\nu \big) \bigg[ \int d\tau' \, D_{\rm ret} (z^\mu, z^{\mu'} ) \bigg]^2  \nonumber \\
	& - \frac{ m^3 c_1^2 c_2 }{ m_{pl}^4 } \big( a^\mu + P^{\mu\nu} \nabla_\nu \big) \int d\tau' d\tau''  \, D_{\rm ret} (z^\mu, z^{\mu'} ) D_{\rm ret} (z^{\mu'}, z^{\mu''} )  ~.
\label{secondforce1}
\end{align}
The $O(V_-^2)$ terms in (\ref{tempS2eff1}) do not contribute to the equations of motion since the variational derivative of those terms with respect to $z_1^\mu$ or $z^\mu_2$ is proportional to $V_- (x; z] \propto z_1^\mu - z_2^\mu$, which vanishes when we set $z_1^\mu = z_2^\mu = z^\mu$ from (\ref{coordend1}).
It will be convenient to evaluate each line of (\ref{secondforce1}) separately, which we do in the next subsections.

\subsubsection{Regularization of $F_{(2a)}^\mu$}

Here, we regularize the divergent integrals appearing in the first line of (\ref{secondforce1}),
\begin{align}
	F^\mu _{(2a)} \equiv - \frac{ m^3 c_1^2 c_2 }{ 2 m_{pl}^4 }  \big( a^\mu + P^{\mu\nu} \nabla_\nu \big) \bigg[ \int d\tau' \, D_{\rm ret} (z^\mu, z^{\mu'} ) \bigg]^2  ~.
\end{align}
The regularized expression for $F^\mu _{(2a)}$ follows simply from substituting (\ref{divintegral3}) and (\ref{divintegral4}) into $F^\mu _{(2a)}$, which gives
\begin{align}
	F^\mu _{(2a)} (\tau) = {} & - \frac{ m^3 c_1^2 c_2}{ m_{pl}^4} \bigg\{
	\frac{\Lambda}{4\pi } \frac{a^\mu}{2}  I_{R} (z^\mu) + \frac{ \Lambda }{ 4\pi} P^{\mu\nu} \nabla_\nu I_{R} (z^\mu)  + \frac{1}{2} \big( a^\mu   + P^{\mu\nu} \nabla_\nu \big) I_{R} (z^\mu)^2 
	 \bigg\}  ~.
\label{secondorder1}
\end{align}

\subsubsection{Regularization of $F_{(2b)}^\mu$}
\label{sec:F2b}

Turn now to regularizing the integrals in the second line of (\ref{secondforce1}),
\begin{align}
	F^\mu _{(2b)} \equiv - \frac{ m^3 c_1^2 c_2 }{ m_{pl}^4 } \big( a^\mu + P^{\mu\nu} \nabla_\nu \big) \int d\tau' d\tau''  \, D_{\rm ret} (z^\mu, z^{\mu'} ) D_{\rm ret} (z^{\mu'}, z^{\mu''} ) ~.
\label{force2b11}
\end{align}
Notice that $F_{(2b)}^\mu$ involves the divergent integral of a divergent integral as opposed to products of divergent integrals as in $F_{(2a)}^\mu$. One way to deal with these integrals is to insert a factor of $1 = \int d\bar{\tau} \, \delta (\bar{\tau} - \tau')$ into (\ref{force2b11}) to ``disentangle'' the $\tau'$ and $\tau''$ integrals,
\begin{align}
	F_{(2b)}^\mu (\tau) = {} & - \frac{ m^3 c_1^2 c_2 }{ m_{pl}^4 } \int d\bar{\tau} \bigg\{ a^\mu \bigg( \int d\tau' \, \delta(\tau' - \bar{\tau}) D_{\rm ret} (z^\mu, z^{\mu'}) \bigg) \bigg( \int d\tau'' \, D_{\rm ret} (z^{\bar{\mu}}, z^{\mu''}) \bigg) \nonumber \\
	& {\hskip1in} +  \bigg( P^{\mu\nu} \int d\tau' \, \delta (\tau' - \bar{\tau}) \nabla_\nu D_{\rm ret} (z^\mu, z^{\mu'}) \bigg) \bigg( \int d\tau'' \, D_{\rm ret} (z^{\bar{\mu} }, z^{\mu''}) \bigg) \bigg\} ~.
\end{align}
The integrals involving the factor of $\delta (\tau' - \bar{\tau})$ can be integrated by letting $\tau' = \tau+s$ and expanding the integrands about $s=0$, as before. Doing this, we find
\begin{align}
	\int d\tau' \, \delta (\tau' - \bar{\tau} )  D_{\rm ret} (z^\mu, z^{\mu'} ) = {} & \frac{\Lambda}{4\pi} \delta ( \tau - \bar{\tau}) +  D_{R} (z^\mu, z^{\bar{\mu}}) 
\label{factorintegral1} \\
	\int d\tau' \, \delta (\tau' - \bar{\tau} )  \nabla_\nu D_{\rm ret} (z^\mu, z^{\mu'} ) = {} &  - \frac{\Lambda}{4\pi} \frac{a_\nu}{2}  \delta ( \tau - \bar{\tau})  + \nabla_\nu D_{R} (z^\mu, z^{\bar{\mu}})  ~.
\label{factorintegral2} 
\end{align}
Inserting these expressions into $F_{(2b)}^\mu$ and integrating over $\bar{\tau}$ then gives
\begin{align}
	F_{(2b)}^\mu (\tau) ={} &  - \frac{ m^3 c_1 ^2 c_2 }{ m_{pl}^4} \bigg\{ \bigg( \frac{\Lambda}{4\pi} \bigg)  \frac{  a^\mu}{2}  \int d\tau'' \, D_{\rm ret} (z^\mu, z^{\mu''}) +  \big( a^\mu + P^{\mu\nu} \nabla_\nu \big) \int d\tau' \, D_{R} (z^\mu, z^{\mu'}) \int d\tau'' \, D_{\rm ret} (z^{\mu'}, z^{\mu''}) \bigg\}  ~.
\end{align}
The remaining integrals over $\tau''$ have been computed already in Section \ref{sec:firstorder}. Substituting in their values from (\ref{divintegral3}), (\ref{divintegral4}) and collecting terms yields
\begin{align}
	F^\mu_{(2b)} (\tau) = {} & - \frac{ m^3 c_1^2 c_2 }{ m_{pl}^4 } \bigg\{ \bigg( \frac{\Lambda}{4\pi} \bigg)^2 \frac{a^\mu}{2} + \bigg( \frac{\Lambda}{4\pi} \bigg) \frac{3 a^\mu}{2} I_{R} (z^\mu) + \bigg( \frac{\Lambda}{4\pi} \bigg) P^{\mu\nu} \nabla_\nu I_R (z^\mu) \nonumber \\
	& {\hskip0.7in} + \big( a^\mu + P^{\mu\nu} \nabla_\nu \big) \int d\tau' \, D_{R} (z^\mu, z^{\mu'}) I_{R} (z^{\mu'}) \bigg\}  ~.
\label{secondorder2}
\end{align}
For the third order calculations below, it is useful to collect an intermediate result that can be inferred from the above manipulations, namely
\begin{align}
	\int d\tau' d\tau'' \, D_{\rm ret} (z^\mu, z^{\mu'}) D_{\rm ret} (z^{\mu'}, z^{\mu''}) = {} & \bigg( \frac{\Lambda}{4\pi} \bigg)^2 + 2 \bigg( \frac{ \Lambda }{ 4\pi} \bigg) I_R (z^\mu) + \int d\tau' \, D_R (z^\mu, z^{\mu'}) I_R (z^{\mu'})  
\label{divintegral5} \\
	P^{\mu\nu} \int d\tau' d\tau'' \, \nabla_\nu D_{\rm ret} (z^\mu, z^{\mu'}) D_{\rm ret} (z^{\mu'}, z^{\mu''})  = {} & - \frac{1}{2} \bigg( \frac{ \Lambda}{4\pi} \bigg) ^2 a^\mu - \frac{ 1}{2} \bigg( \frac{ \Lambda}{4\pi} \bigg) a^\mu I_R (z^\mu)  + \bigg( \frac{ \Lambda}{ 4\pi} \bigg) P^{\mu\nu} \nabla_\nu I_R( z^\mu) \nonumber \\
	& + P^{\mu\nu} \nabla_\nu \int d\tau' \, D_R (z^\mu, z^{\mu'}) I_R (z^{\mu'})  ~.
\label{divintegral6}
\end{align}

\subsection{Self force at third order}

\begin{figure}
	\includegraphics[width=9cm]{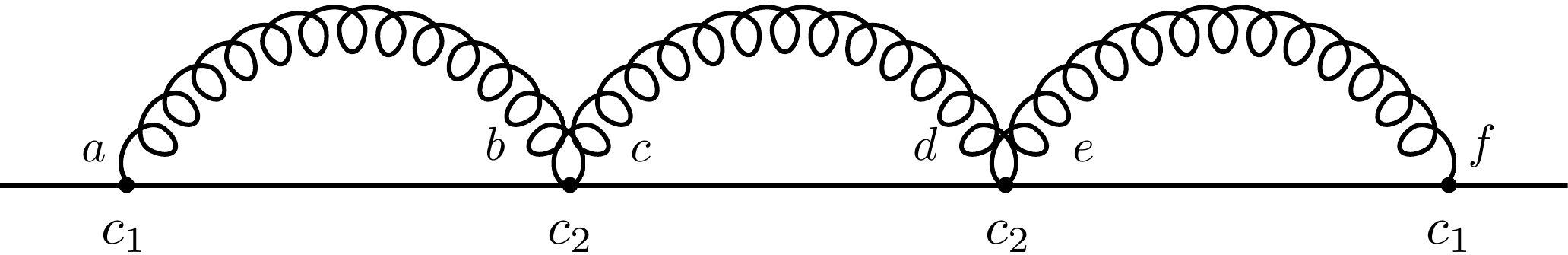} \\ ~\\
	\includegraphics[width=6.5cm]{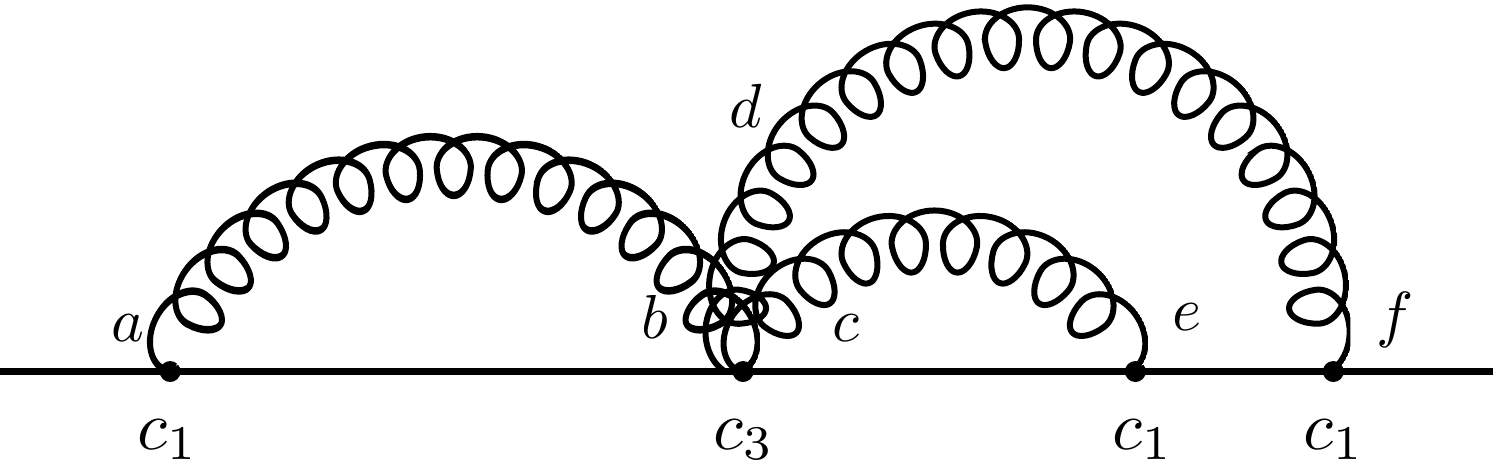}
\caption{The diagrams contributing to the effective action at third order in $\varepsilon$. Power counting indicates that these diagrams each scale as $\sim \varepsilon^3 L$.}
\label{fig:thirdorder}
\end{figure}

At third order in $\varepsilon$ there are two diagrams contributing to the effective action and are shown in Figure \ref{fig:thirdorder}. The Feynman rules imply that the corresponding effective action is
\begin{align}
	S^{(3)}_{\rm eff}  [ z_{1,2}^\mu ]  = {} & \frac{ 1}{2!} \int_x \int_{x'} \int_{x''} \int_{x'''} T_a (x; z] D^{ab} (x,x') T_{bc} (x') D^{cd} (x', x'') T_{de} (x'') D^{ef} (x'',x''') T_f (x''')  \\
	& + \frac{1}{3!} \int_x \int _{x'} \int_{x''} \int_{x'''} T_a(x; z] D^{ab} (x,x') T_{bcd} (x') D^{ce}(x', x'') T_e (x'') D^{df} (x', x''') T_f (x''')  ~.
\end{align}
In the $\pm$ basis, the nonzero elements of the $c_{bcdg}$ tensor, which appears in the expression for the $T_{bcd}$ vertex, are
\begin{align}
	1 & = c_{---+} = c_{--+-} = c_{-+--} = c_{+---} \\
	1/4 & = c_{-+++} = c_{+-++} = c_{++-+} = c_{+++-}
\end{align}
so that  the part of the effective action that contributes to the worldline equations of motion is
\begin{align}
	S^{(3)}_{\rm eff} [ z_{1,2}^\mu ] = {} & \frac{ m^4 c_1^2 c_2^2}{ m_{pl}^6 } \int_x \int_{x'} \int_{x''} \int_{x'''} \bigg\{ V_- (x; z] D_{\rm ret} (x,x') V_+ (x'; z] D_{\rm ret} (x', x'') V_+ (x''; z] D_{\rm ret} (x'', x''') V_+ (x'''; z] \nonumber \\
	& {\hskip1in} + V_- (x; z] D_{\rm ret} (x,x') V_+ (x'; z] D_{\rm ret} (x', x'') V_+ (x''; z] D_{\rm ret} (x, x''') V_+ (x'''; z] + O(V_-^2) \bigg\} \nonumber \\
	& + \frac{ m^4 c_1^3 c_3 }{ 6 m_{pl}^6 } \int_x \int_{x'} \int_{x''} \int_{x'''} \bigg\{ 3 V_- (x; z] D_{\rm ret} (x,x') V_+ (x') D_{\rm ret} (x', x'') V_+ (x''; z] D_{\rm ret} (x', x''') V_+ (x'''; z] \nonumber \\
	& {\hskip1in} + V_- (x; z] D_{\rm ret} (x, x') V_+ (x'; z] D_{\rm ret} (x, x'') V_+ (x''; z] D_{\rm ret} (x,x''') V_+ (x''' ; z] + O(V_-^2) \bigg\}  ~.
\label{thirdeffaction1}
\end{align}
Variation of (\ref{thirdeffaction1}) gives the contribution to the equations of motion from the third order self force,
\begin{align}
	F^\mu _{(3)} = {} & \frac{ m^4 c_1^2 c_2^2 }{ m_{pl}^6 } \big( a^\mu + P^{\mu\nu} \nabla_\nu \big) \int d\tau' d\tau'' d\tau'''  \, D_{\rm ret} (z^\mu, z^{\mu'}) D_{\rm ret} (z^{\mu'}, z^{\mu''} ) D_{\rm ret} (z^{\mu''}, z^{\mu'''}) \nonumber \\
	& + \frac{ m^4 c_1^2 c_2^2 }{ m_{pl}^6 } \big( a^\mu + P^{\mu\nu} \nabla_\nu \big) \int d\tau' d\tau''' \,  D_{\rm ret} (z^\mu, z^{\mu'}) D_{\rm ret} (z^\mu, z^{\mu'''}) \int d\tau'' \, D_{\rm ret} (z^{\mu'}, z^{\mu''}) \nonumber \\
	& + \frac{ m^4 c_1^3 c_3 }{ 2 m_{pl}^6 } \big( a^\mu + P^{\mu\nu} \nabla_\nu \big) \int d\tau' \, D_{\rm ret} (z^\mu, z^{\mu'} ) \bigg[ \int d\tau'' \, D_{\rm ret} (z^{\mu'} , z^{\mu''} ) \bigg]^2 \nonumber \\
	& + \frac{ m^4 c_1^3 c_3 }{ 6 m_{pl}^6} \big( a^\mu + P^{\mu\nu} \nabla_\nu \big) \bigg[ \int d\tau' \, D_{\rm ret} (z^\mu, z^{\mu'}) \bigg]^3 \label{thirdorder0} \\
	= {} & F^\mu _{(3a)} (\tau) + F^\mu _{(3b)} (\tau) + F^\mu _{(3c)} (\tau) + F^\mu _{(3d)} (\tau)  ~.
\end{align}
We next regularize each line appearing in $F^\mu _{(3)}$. Only two new integrals need to be regularized and both appear  in $F^\mu _{(3a)}$. The remaining integrals in (\ref{thirdorder0}) can be expressed in terms of the previously regularized integrals in (\ref{divintegral3}), (\ref{divintegral4}), (\ref{divintegral5}) and (\ref{divintegral6}).

\subsubsection{Regularization of $F_{(3a)}^\mu$}

Regularizing $F^\mu _{(3a)}$ follows similar manipulations and steps as in regularizing $F^\mu_{(2b)}$ from Section \ref{sec:F2b} in that we ``factorize'' the proper time integrals by inserting $1 = \int d\bar{\tau} \delta (\tau' - \bar{\tau})$,
\begin{align}
	F^\mu _{(3a)} \equiv {} &  \frac{ m^4 c_1^2 c_2^2 }{ m_{pl}^6} \int d\bar{\tau} \bigg\{ a^\mu \bigg( \int d\tau' \, \delta ( \tau' - \bar{\tau} ) D_{\rm ret} (z^\mu, z^{\mu'}) \bigg) \bigg( \int d\tau'' d\tau''' \, D_{\rm ret} (z^{\bar{\mu}}, z^{\mu''}) D_{\rm ret} (z^{\mu''}, z^{\mu'''}) \bigg) \nonumber \\
	& {\hskip0.5in} + \bigg( P^{\mu\nu} \int d\tau' \, \delta ( \tau' - \bar{\tau} ) \nabla_\nu D_{\rm ret} (z^\mu, z^{\mu'}) \bigg) \bigg( \int d\tau'' d\tau''' \, D_{\rm ret} (z^{\bar{\mu}}, z^{\mu''}) D_{\rm ret} (z^{\mu''} , z^{\mu'''}) \bigg)
	\bigg\}  ~.
\end{align}
Using (\ref{factorintegral1}) and (\ref{factorintegral2}) to regularize the $\tau'$ integrals gives
\begin{align}
	F^\mu _{(3a)} = {} &  \frac{ m^4 c_1^2 c_2^2 }{ m_{pl}^6} \bigg\{ \bigg( \frac{\Lambda}{4\pi} \bigg) \frac{a^\mu }{ 2} \int d\tau'' d\tau''' \, D_{\rm ret} (z^{\mu}, z^{\mu''}) D_{\rm ret} (z^{\mu''}, z^{\mu'''}) \nonumber \\
		& {\hskip0.9in} + \big( a^\mu + P^{\mu\nu} \nabla_\nu \big) \int d\tau' \, D_R (z^\mu, z^{\mu'}) \int d\tau'' d\tau''' \, D_{\rm ret} (z^{\mu'}, z^{\mu''}) D_{\rm ret} (z^{\mu''}, z^{\mu'''})  ~.
\end{align}
The remaining integrals over $\tau''$ and $\tau'''$ are given by (\ref{divintegral5}). Substitution then gives
\begin{align}
	F^\mu _{(3a)} = {} & \frac{ m^4 c_1^2 c_2^2 }{ m_{pl}^6 } \bigg\{ \bigg( \frac{ \Lambda }{ 4\pi} \bigg)^3 \frac{ a^\mu }{ 2}  + 2  \bigg( \frac{ \Lambda }{ 4\pi} \bigg)^2 a^\mu I_R (z^\mu) +  \bigg( \frac{ \Lambda }{ 4\pi} \bigg)^2 P^{\mu\nu} \nabla_\nu I_R (z^\mu) + \frac{5}{2}  \bigg( \frac{ \Lambda }{ 4\pi} \bigg) \int d\tau' \, D_R (z^\mu, z^{\mu'}) I_R (z^{\mu'}) \nonumber \\
	& {\hskip0.25in} + 2  \bigg( \frac{ \Lambda }{ 4\pi} \bigg) P^{\mu\nu} \nabla_\nu \int d\tau' \, D_R (z^\mu, z^{\mu'}) I_R (z^{\mu'}) + \big( a^\mu + P^{\mu\nu} \nabla_\nu \big) \int d\tau' d\tau'' \, D_R (z^\mu, z^{\mu'}) D_R (z^{\mu'}, z^{\mu''}) I_R (z^{\mu''}) \bigg\}  ~.
\label{thirdorder1}
\end{align}

\subsubsection{Regularization of $F_{(3b)}^\mu$}

Isolating the singular pieces in $F^\mu _{(3b)}$ turns out to be relatively simple since some of the integrals already factor giving
\begin{align}
	F^\mu _{(3b)} \equiv {} & \frac{ m^4 c_1^2 c_2^2}{ m_{pl}^6 } \bigg\{ a^\mu \bigg( \int d\tau' d\tau'' \, D_{\rm ret} (z^\mu, z^{\mu'}) D_{\rm ret} (z^{\mu'}, z^{\mu''}) \bigg) \bigg( \int d\tau''' \, D_{\rm ret} (z^\mu, z^{\mu'''}) \bigg) \nonumber \\
		& {\hskip0.5in} + \bigg( P^{\mu\nu} \int d\tau' d\tau'' \, \nabla_\nu D_{\rm ret} (z^\mu, z^{\mu'}) D_{\rm ret} (z^{\mu'}, z^{\mu''}) \bigg) \bigg( \int d\tau''' \, D_{\rm ret} (z^\mu, z^{\mu'''}) \bigg) \nonumber \\
		& {\hskip0.5in} + \bigg( \int d\tau' d\tau'' \, D_{\rm ret} (z^\mu, z^{\mu'}) D_{\rm ret} (z^{\mu'}, z^{\mu''}) \bigg) \bigg( P^{\mu\nu} \int d\tau''' \, \nabla_\nu D_{\rm ret} (z^\mu, z^{\mu'''}) \bigg)  \bigg\}  ~.
\end{align}
Substituting the expressions (\ref{divintegral3}), (\ref{divintegral4}), (\ref{divintegral5}) and (\ref{divintegral6}) into $F^\mu_{(3b)}$ then gives
\begin{align}
	F^\mu _{(3b)} = {} & \frac{ m^4 c_1^2 c_2^2}{ m_{pl}^6} \bigg\{ \bigg( \frac{ \Lambda}{ 4\pi} \bigg)^2 a^\mu I_R (z^\mu) + 2 \bigg( \frac{ \Lambda}{ 4\pi} \bigg)^2 P^{\mu\nu} \nabla_\nu I_R (z^\mu) + \frac{3}{2} \bigg( \frac{ \Lambda}{ 4\pi} \bigg) I_R^2 (z^\mu) + 3 \bigg( \frac{ \Lambda}{ 4\pi} \bigg) P^{\mu\nu} \nabla_\nu I_R (z^\mu ) I_R (z^\mu) \nonumber \\
	& {\hskip0.25in} + \frac{1}{2} \bigg( \frac{ \Lambda}{ 4\pi} \bigg) \int d\tau' \, D_R (z^\mu, z^{\mu'}) I_R (z^{\mu'}) + \bigg( \frac{ \Lambda}{ 4\pi} \bigg) P^{\mu\nu} \nabla_\nu \int d\tau' \, D_R (z^\mu, z^{\mu'}) I_R (z^{\mu'}) \nonumber \\
	& {\hskip0.25in} + I_R (z^\mu ) \big( a^\mu + P^{\mu\nu} \nabla_\nu \big) \int d\tau' \, D_R (z^\mu, z^{\mu'}) I_R (z^{\mu'}) + P^{\mu\nu} \nabla_\nu I_R (z^\mu) \int d\tau' \, D_R (z^\mu, z^{\mu'}) I_R(z^{\mu'}) \bigg\}  ~.
\label{thirdorder2}
\end{align}

\subsubsection{Regularization of $F_{(3c)}^\mu$ and $F_{(3d)}^\mu$}

Evaluating the singular pieces of $F^\mu _{(3c)}$ and $F^\mu _{(3d)}$ is simply performed by substituting in (\ref{divintegral3}) and (\ref{divintegral4}) and expanding to find
\begin{align}
	F^\mu_{(3c)}+F^\mu_{(3d)} \equiv {} & \frac{ m^4 c_1^3 c_3 }{2 m_{pl}^6} \bigg\{ \bigg( \frac{\Lambda}{4\pi} \bigg)^3 \frac{a^\mu}{3} + 2  \bigg( \frac{\Lambda}{4\pi} \bigg)^2 \big( a^\mu  + P^{\mu\nu} \nabla_\nu  \big)  I_R (z^\mu)+  \bigg( \frac{\Lambda}{4\pi} \bigg) \big( a^\mu   + P^{\mu\nu} \nabla_\nu  \big) I_R^2 (z^\mu) \nonumber \\
	&{\hskip0.55in} + 2  \bigg( \frac{\Lambda}{4\pi} \bigg) \big( a^\mu + P^{\mu\nu} \nabla_\nu \big) \int d\tau' \, D_R (z^\mu, z^{\mu'}) I_R(z^{\mu'}) \nonumber \\
	& {\hskip0.55in} + \frac{ 1}{3} a^\mu I_R^3 (z^\mu) + \big( a^\mu + P^{\mu\nu} \nabla_\nu \big) \int d\tau' \, D_R (z^\mu, z^{\mu'}) I_R (z^{\mu'}) \bigg\}  ~.
\label{thirdorder3}
\end{align}

\section{Renormalization}
\label{sec:renormalization}

\begin{figure}
	\includegraphics[width=4.5cm]{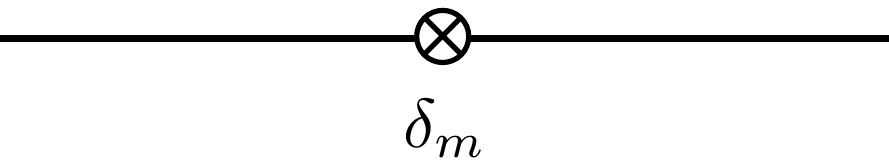} {\hskip0.25in} \\ ~\\
	\includegraphics[width=4.5cm]{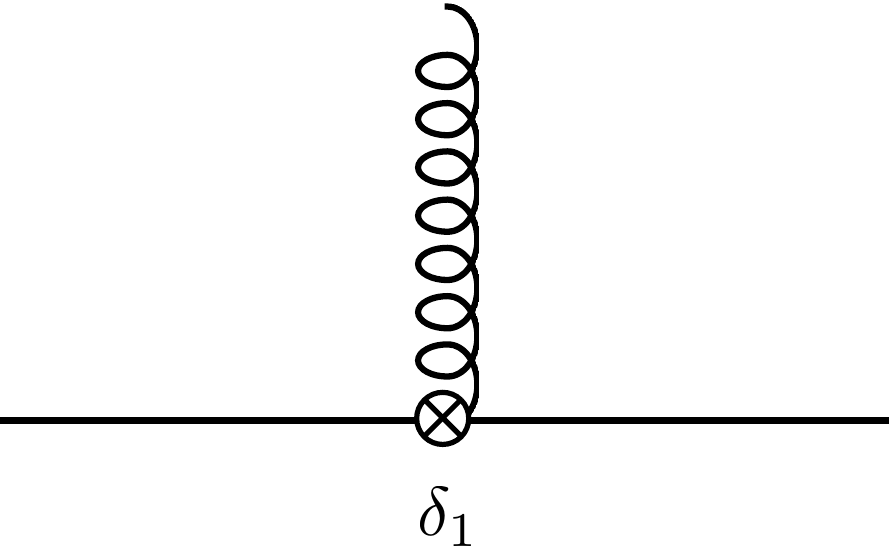} {\hskip0.25in}
	\includegraphics[width=4.5cm]{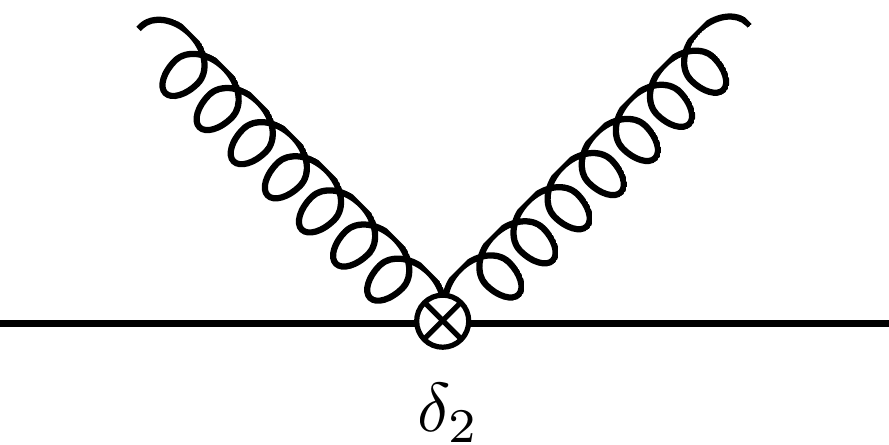}
\caption{Feynman rules for the counter terms that cancel the divergences in the self force through third order in $\varepsilon$. Here, $\delta_m$ is a mass counter term while $\delta_1$ and $\delta_2$ can be associated with renormalizing $c_1$ and $c_2$.}
\label{fig:counterterms}
\end{figure}

Combining our results for the first, second and third order self force corrections from (\ref{firstorder7}), (\ref{secondorder1}), (\ref{secondorder2}), (\ref{thirdorder1}), (\ref{thirdorder2}) and (\ref{thirdorder3}) gives the regularized self force through third order in $\varepsilon$,
\begin{align}
	F^\mu (\tau) = {} & a^\mu \bigg\{ \frac{ m^2 c_1^2 }{ 2 m_{pl}^2 } \bigg( \frac{ \Lambda }{ 4\pi } \bigg) - \frac{ m^3 c_1^2 c_2 }{ 2 m_{pl}^4 } \bigg( \frac{ \Lambda }{ 4\pi } \bigg)^2 + \frac{ m^4 c_1^2 c_2^2 }{ 2 m_{pl}^6 } \bigg( \frac{ \Lambda }{ 4\pi } \bigg)^3 + \frac{ m^4 c_1^3 c_3 }{ 6 m_{pl}^6 } \bigg( \frac{ \Lambda }{ 4\pi } \bigg)^3 \bigg\} \nonumber \\
	& + \big( a^\mu + P^{\mu\nu} \nabla_\nu \big) I_R(z^\mu)  \bigg\{ \frac{ m^2 c_1^2 }{ m_{pl}^2 } - \frac{ 2 m^3 c_1^2 c_2 }{ m_{pl}^4 } \bigg( \frac{ \Lambda }{ 4\pi } \bigg) + \frac{ 3 m^4 c_1^2 c_2^2 }{ m_{pl}^6} \bigg( \frac{ \Lambda }{ 4\pi } \bigg)^2 + \frac{ m^4 c_1^3 c_3 }{ m_{pl}^6 } \bigg( \frac{ \Lambda }{ 4\pi } \bigg)^2 \bigg\} \nonumber \\
	& + \big(a^\mu  + P^{\mu\nu} \nabla_\nu \big) \bigg( \frac{1}{2}  I_R^2 (z^\mu) +\int d\tau' \, D_R (z^\mu, z^{\mu'}) I_R (z^{\mu'}) \bigg) \bigg\{ - \frac{ m^3 c_1^2 c_2}{m_{pl}^4 } + \frac{ 3 m^4 c_1^2 c_2^2 }{ m_{pl}^6} \bigg( \frac{ \Lambda }{ 4\pi } \bigg) + \frac{ m^4 c_1^3 c_3 }{ m_{pl}^6} \bigg( \frac{ \Lambda }{ 4\pi } \bigg) \bigg\} \nonumber \\
	& + \frac{ m^4 c_1^2 c_2^2}{ m_{pl}^6}  \big( a^\mu + P^{\mu\nu} \nabla_\nu \big) \bigg( I_R (z^\mu)   \int d\tau' \, D_R (z^{\mu}, z^{\mu'}) I_R ( z^{\mu'}) + \int d\tau' d\tau''  \, D_R (z^\mu, z^{\mu'}) D_R (z^{\mu'}, z^{\mu''}) I_R (z^{\mu''}) \bigg) \nonumber \\
	& + \frac{ m^4 c_1^3 c_3}{ 2 m_{pl}^6} \big( a^\mu + P^{\mu\nu} \nabla_\nu \big) \bigg( \frac{ 1}{3}  I_R^3 (z^\mu)  + \int d\tau' \, D_R (z^{\mu}, z^{\mu'} ) I_R^2 (z^{\mu'}) \bigg) + O(\varepsilon^4)  ~.
\label{regselfforce1}
\end{align}
The appearance of higher-order divergences (e.g., $\sim \Lambda^3 a^\mu$) and history-dependent divergences (e.g., $\sim \Lambda P^{\mu\nu} \nabla_\nu I_R^2(z^\mu)$) may seem unsettling. However, despite this seemingly pathological structure all the divergences through third order in $\varepsilon$ can be absorbed into purely {\it local} and {\it time-independent} counter terms in the effective action while the remaining terms independent of $\Lambda$ cannot.

\begin{figure}
	\includegraphics[width=4.5cm]{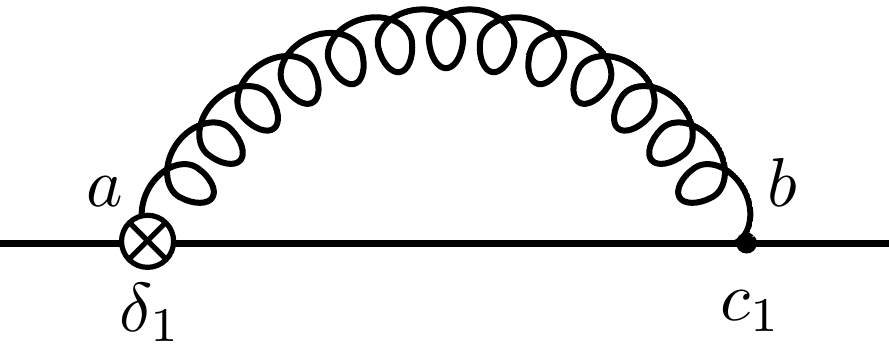} {\hskip0.25in}
	\includegraphics[width=4.5cm]{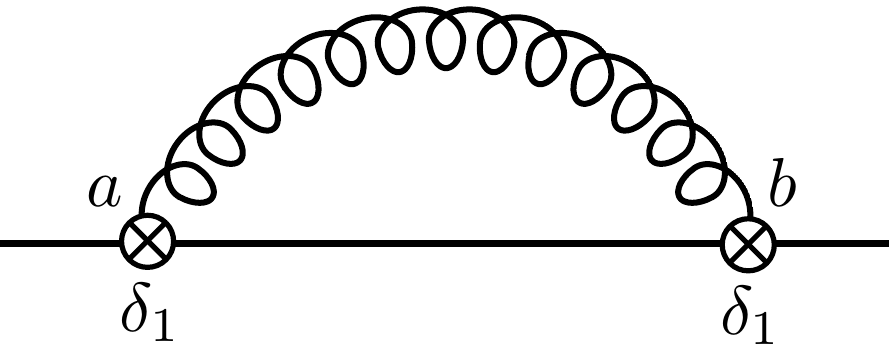} \\ ~\\
	\includegraphics[width=7.5cm]{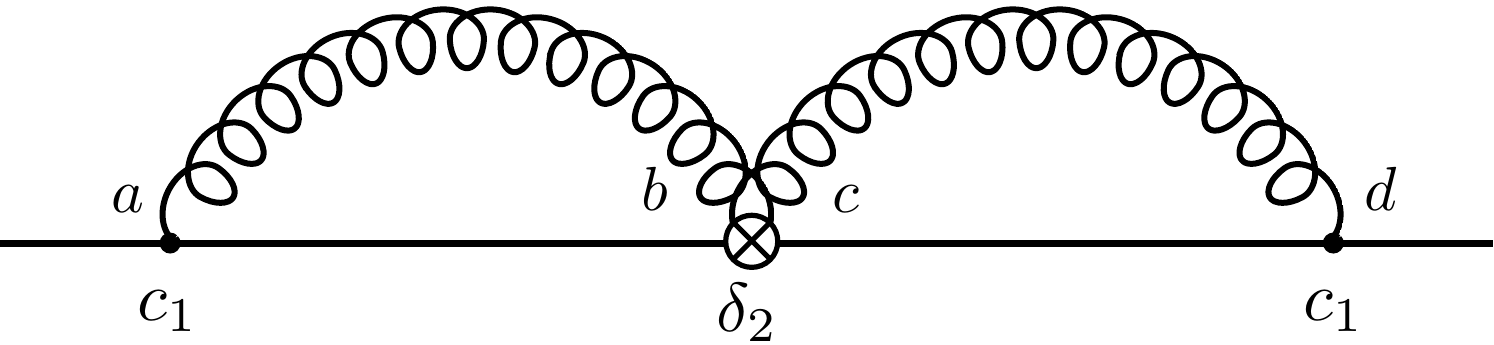} {\hskip0.25in}
	\includegraphics[width=7.5cm]{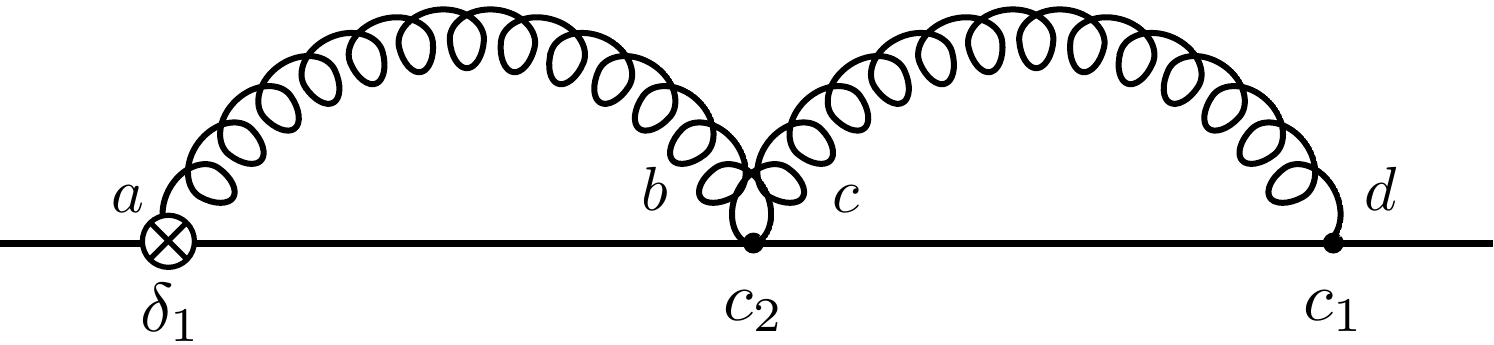}
\caption{The counter term diagrams that contribute to renormalizing the self force through third order in $\varepsilon$.}
\label{fig:sfcounterterms}
\end{figure}

More specifically, we add to the action in (\ref{nonlinear2}) the following counter terms
\begin{align}
	- \delta_m \int d\tau - \frac{ \delta_1 }{ m_{pl}} \int d\tau \, \psi (z^\mu) - \frac{ \delta_2 }{ 2 m_{pl}^2 } \int d\tau \, \psi^2 (z^\mu)  ~.
\label{counterterms1}
\end{align}
The Feynman rules for the counter terms above are shown in Figure \ref{fig:counterterms} and their contributions to the effective action (and hence the self force) are given by the Feynman diagrams in Figure \ref{fig:sfcounterterms}. Following similar steps we used in deriving and regularizing the self force through third order above one can show that the contributions of the counter terms to the self force are
\begin{align}
	F^\mu _{CT} = {} & a^\mu \bigg\{ - \delta_m + \frac{ m c_1 \delta_1 }{ m_{pl}^2 } \bigg( \frac{ \Lambda }{ 4\pi } \bigg) - \frac{ m^2 c_1^2 \delta_2 }{ 2 m_{pl}^4 }  \bigg( \frac{ \Lambda }{ 4\pi } \bigg)^2 - \frac{ m^2 c_1 c_2 \delta_1 }{ m_{pl}^4 } \bigg( \frac{ \Lambda }{ 4\pi } \bigg)^2 + \frac{ \delta_1^2 }{ 2 m_{pl}^2 } \bigg( \frac{ \Lambda }{ 4\pi } \bigg)  \bigg\} \nonumber \\
	& + \big( a^\mu + P^{\mu\nu} \nabla_\nu \big) I_R(z^\mu) \bigg\{ \frac{ 2 m c_1 \delta_1}{ m_{pl}^2 } - \frac{ 2 m^2 c_1^2 \delta_2 }{ m_{pl}^4 } \bigg( \frac{ \Lambda }{ 4\pi } \bigg) - \frac{ 4 m^2 c_1 c_2 \delta_1 }{ m_{pl}^4 } \bigg( \frac{ \Lambda }{ 4\pi } \bigg) + \frac{ \delta_1^2 }{ m_{pl}^2 } \bigg\} \nonumber \\
	& + \big( a^\mu + P^{\mu\nu} \nabla_\nu \big) \bigg( \frac{1}{2}  I_R^2 (z^\mu)  + \int d\tau' \, D_R (z^\mu, z^{\mu'}) I_R (z^{\mu'}) \bigg) \bigg\{ - \frac{ m^2 c_1^2 \delta_2 }{ m_{pl}^4} - \frac{ 2 m^2 c_1 c_2 \delta_1 }{ m_{pl}^4 } \bigg\} + O(\varepsilon^4)  ~.
\label{CTforce1}
\end{align}
Notice that the divergent terms automatically organize themselves to have the same structure as the divergent terms appearing in (\ref{regselfforce1}). That is, the divergences appearing in both (\ref{regselfforce1}) and (\ref{CTforce1}) are proportional to $a^\mu$, $(a^\mu + P^{\mu\nu} \nabla_\nu) I_R$ and $ (a^\mu + P^{\mu\nu} \nabla_\nu) ( I_R^2 / 2 + \cdots )$. This is no accident. In order for the divergences to be removed, the structure of the counter terms must exactly match the divergent parts of the self force. This can be viewed as a consistency check that our calculations to this point are correct. In addition, the terms in (\ref{regselfforce1}) not involving $\Lambda$ cannot be absorbed by these counter terms.

The divergences in $F^\mu (\tau)$ are cancelled by requiring $\delta_m$, $\delta_1$ and $\delta_2$ to eliminate all terms proportional to $\Lambda$, which implies
\begin{align}
	\delta_m = {} & \frac{ m^2 c_1^2 }{ 2 m_{pl}^2 } \bigg( \frac{ \Lambda }{ 4\pi } \bigg) + \frac{ m^3 c_1^2 c_2 }{ 2m_{pl}^4 } \bigg( \frac{ \Lambda }{ 4\pi } \bigg)^2 + \frac{ m^4 c_1^2 c_2^2 }{ 2 m_{pl}^6 } \bigg( \frac{ \Lambda }{ 4\pi } \bigg)^3 + \frac{ m^4 c_1^3 c_3 }{ 6 m_{pl}^6 } \bigg( \frac{ \Lambda }{ 4\pi } \bigg)^3 
\label{counterterms10} \\
	\delta_1 = {} &  \frac{ m^2 c_1 c_2}{ m_{pl}^2 } \bigg( \frac{ \Lambda }{ 4\pi } \bigg) + \frac{ m^3 c_1 c_2^2 }{ m_{pl}^4 } \bigg( \frac{ \Lambda }{ 4\pi } \bigg)^2 + \frac{ m^3 c_1^2 c_3 }{ 2 m_{pl}^4} \bigg( \frac{ \Lambda }{ 4\pi } \bigg)^2 
\label{counterterms11} \\
	\delta_2 = {} & \frac{ m^2 c_2^2 }{ m_{pl}^2 } \bigg( \frac{ \Lambda }{ 4\pi } \bigg) + \frac{ m^2 c_1 c_3}{ m_{pl}^2 } \bigg( \frac{ \Lambda }{ 4\pi } \bigg)  ~.
\label{counterterms12}
\end{align}
Notice that there are three divergent terms appearing in the self force in (\ref{regselfforce1}) and there are three counter terms that absorb these divergences by renormalizing $m$, $c_1$ and $c_2$.
Had we computed the self force only up through second order in $\varepsilon$ then (\ref{regselfforce1}) would only have two divergent terms, proportional to $a^\mu$ and $(a^\mu + P^{\mu\nu} \nabla_\nu) I_R$, so that only two counter terms would need to be introduced ($\delta_m$ and $\delta_1$). The key point is that at $O(\varepsilon^n$) the self force will have $n$ divergent terms that can be removed by introducing $n$ counter terms into the action so that the theory can be made finite at any order. Therefore, {\it the nonlinear scalar model is renormalizable at any given order of} $\varepsilon$. This statement is independent of which field variables are used ($\phi$ or $\psi$), since the actions are the same in either case.

With all divergences cancelled by the appropriate counter terms, the renormalized self force through third order in $\varepsilon$ is given by
\begin{align}
	F^\mu (\tau) = {} &  \big( a^\mu + P^{\mu\nu} \nabla_\nu \big) \bigg\{ \frac{ m^2 c_1^2 }{ m_{pl}^2 } I_R( z^\mu)  - \frac{ m^3 c_1^2 c_2}{m_{pl}^4 } \bigg( \frac{1}{2} I_R^2 (z^\mu)  + \int d\tau' \, D_R (z^\mu, z^{\mu'}) I_R (z^{\mu'}) \bigg) \nonumber \\
	& + \frac{ m^4 c_1^2 c_2^2}{ m_{pl}^6} \bigg( I_R (z^\mu) \int d\tau' \, D_R (z^{\mu}, z^{\mu'}) I_R ( z^{\mu'}) + \int d\tau' d\tau''  \, D_R (z^\mu, z^{\mu'}) D_R (z^{\mu'}, z^{\mu''}) I_R (z^{\mu''}) \bigg) \nonumber \\
	& + \frac{ m^4 c_1^3 c_3}{ 2 m_{pl}^6} \bigg( \frac{ 1}{3}  I_R^3 (z^\mu) + \int d\tau' \, D_R (z^{\mu}, z^{\mu'} ) I_R^2 (z^{\mu'}) \bigg) +O(\varepsilon^4) \bigg\}  ~,
\label{renormalizedsf1}
\end{align}
which is the main result of this paper.

As stated before, in dimensional regularization all power-divergent integrals vanish so that $\Lambda = 0$. Therefore, whenever a regularized integral gives a term proportional to $\Lambda$ we can simply drop that term from the computation. This is not the case with log-divergent integrals, which describe physically relevant screening effects. Instead, one must deal with these divergences as we have here by identifying the counter terms that cancel them. However, logarithmic divergences depend on the renormalization scale and induce a classical renormalization group (RG) running of the appropriate coupling constant. The possibility for classical RG running was first recognized in Ref.\,\cite{GoldbergerRothstein:PRD73}. 

In the nonlinear scalar model presented here, 
all interactions occur between the field and the particle only (i.e., there are no field self-interactions). Hence, the only kinds of divergences that can be generated are proportional to $\Lambda^p$ for some integer $p>0$, which vanishes in dimensional regularization. In other words, the theory described by the action in (\ref{nonlinear2}), or equivalently in (\ref{nonlinear1}), {\it does not generate any logarithmically divergent integrals, only power-divergent ones}. Therefore, in this model we can simply make the replacement
\begin{align}
	D_{\rm ret} (z^\mu, z^{\mu'} ) \to D_R ( z^\mu, z^{\mu'} )
\label{replacement1}
\end{align}
wherever we see a retarded propagator in an expression connecting two points on the worldline -- the resulting expression is regular. 

Dimensional regularization is certainly a convenient tool for regularizing the self force in the nonlinear scalar model. However, its use is necessary for gauge theories, e.g., for the gravitational self force. The reason for this is simply that the counter terms induced by the divergences are not necessarily restricted to be gauge invariant under small coordinate transformations on the background spacetime. 
For example, 
in calculating the finite-size corrections to the radiation-reaction force on an electrically charged extended body one must include counter terms such as $\int d\tau \, A_\mu A^\mu$ and $\int d\tau \, a^\mu A_\mu$, among others, which are obviously not gauge invariant \cite{GalleyLeibovichRothstein:PRL105}.
Thus, using a cutoff regularization prescription, for example, would render the resulting theory finite but gauge violating. In dimensional regularization, the counter terms are zero for power-divergent integrals and the renormalized theory is finite {\it and} remains gauge invariant.
The fact that dimensional regularization preserves the symmetries of the theory \cite{tHooftVeltman:NuclPhysB44} is then quite crucial for computing the regular part of higher-order gravitational self force corrections.

We could have obtained (\ref{renormalizedsf1}) with much less work had we simply used the replacement rule (\ref{replacement1}) in the formally divergent self force expression in (\ref{regselfforce1}). However, it is important to highlight the consistency of our renormalization program, especially since one may worry about the consistency of a point particle treatment for calculating higher-order self force corrections and/or about the subtleties associated with renormalizing gauge theories at higher orders, as discussed above in the previous paragraph.

\section{Equations of motion through third order}
\label{sec:eom}

The regular self force equations of motion are given by $ma^\mu = F^\mu$ where $F^\mu$ is given through third order in $\varepsilon$ by (\ref{renormalizedsf1}). All of the terms proportional to $a^\mu$ in this equation can be combined into one function $\Gamma_{\rm DW} (z^\mu)$ so that the effective mass of the compact object is $m_{\rm eff} (\tau) = m \Gamma_{\rm DW} (z^\mu)$ where
\begin{align}
	\Gamma_{\rm DW} (z^\mu) = {} & 1 - \frac{ m c_1^2 }{ m_{pl}^2} I_R (z^\mu) + \frac{ m^2 c_1^2 c_2 }{  m_{pl}^4 } \bigg( \frac{1}{2} I_R^2 (z^\mu) +\int d\tau' \, D_R (z^\mu, z^{\mu'}) I_R (z^{\mu'}) \bigg) \nonumber \\
	& - \frac{ m^3 c_1^2 c_2^2}{m_{pl}^6 } \bigg( I_R (z^\mu) \int d\tau' \, D_R (z^\mu, z^{\mu'}) I_R (z^{\mu'}) + \int d\tau' d\tau'' \, D_R (z^\mu, z^{\mu'} ) D_R (z^{\mu'}, z^{\mu''}) I_R (z^{\mu''}) \bigg) \nonumber \\
	& - \frac{ m^3 c_1^3 c_3 }{ 2 m_{pl}^6} \bigg( \frac{1}{3} I_R ^3 (z^\mu) + \int d\tau' \, D_R (z^\mu, z^{\mu'}) I_R^2 (z^{\mu'}) \bigg) + O(\varepsilon^4)  ~.
\label{effmass3} 
\end{align}
In doing so, we find that the equations of motion can be written as
\begin{align}
	m \Gamma_{\rm DW} (z^\mu) a^\mu = - m P^{\mu\nu} \nabla_\nu \Gamma_{\rm DW} (z^\mu)
\label{renormalizedsf2}
\end{align}
or, equivalently, dividing both sides by $m \Gamma_{\rm DW}$
\begin{align}
	a^\mu = - P^{\mu\nu} \nabla_\nu \ln \Gamma_{\rm DW} (z^\mu)  ~.
\label{renormalizedsf3}
\end{align}
The form of (\ref{renormalizedsf3}) suggests that $\Gamma_{\rm DW} (z^\mu ) = C ( \psi_R / m_{pl})$ when comparing with (\ref{sfeom2}). If this is true, then the regular part of $\psi$ in the Detweiler-Whiting scheme is inferred to be
\begin{align}
	\frac{\psi_R (x) }{ m_{pl} } = {} & - \frac{ m c_1}{ m_{pl}^2 } I_R (x) + \frac{ m^2 c_1 c_2}{ m_{pl}^4 } \int d\tau \, D_R (x, z^{\mu}) I_R (z^{\mu}) - \frac{ m^3 c_1 c_2^2}{ m_{pl}^6} \int d\tau d\tau' \, D_R (x, z^\mu) D_R(z^\mu, z^{\mu'}) I_R (z^{\mu'}) \nonumber \\
	& - \frac{ m^3 c_1^2 c_3}{ 2 m_{pl}^6} \int d\tau \, D_R(x, z^\mu) I_R^2 (z^\mu)  + O(\varepsilon^4)~.
\label{psiR1}
\end{align}
In the next paper in this series we will show that (\ref{psiR1}) is indeed the correct expression for the $R$-part of $\psi$ through $O(\varepsilon^3)$ by explicitly calculating the radiative field emitted by the SCO through third order \cite{Galley:Nonlinear2}. We will also compute the self force through third order from these emitted waves and find agreement with (\ref{renormalizedsf3}), thereby establishing a complete formalism that can self-consistently compute the SCO equations of motion and waveform.

One can interpret $m \Gamma_{\rm DW} (z^\mu)$ as an effective mass as in (\ref{renormalizedsf2}) or equivalently as an alteration to the self force experienced by the SCO as in (\ref{renormalizedsf3}). If the effective mass of the SCO, $m \Gamma_{\rm DW}$, decreases under a constant self force, then the body is accelerated to higher velocities because there is effectively less inertia to resist the influence of the force. On the other hand, one could interpret the same dynamics as arising from a body with constant inertia but with a stronger self force acting on the body compared to the former case, which likewise makes the body move faster. The points of view (and hence where we place $\Gamma_{\rm DW} (z^\mu)$ in the equations of motion) give rise to the same dynamics despite the ambiguity in the interpretation. It is likely that such an ambiguity is also manifest in higher order gravitational self force corrections and probably originates from trying to localize the mass-energy of the SCO interacting with the background spacetime.

Instead of splitting the retarded propagator into its regular and singular parts we could have equally chosen to decompose $D_{\rm ret}(x,x')$ into its direct and tail parts as in Section \ref{sec:linearsf}. We will not reproduce the steps for deriving and renormalizing the self force through third order in the latter decomposition since they are similar to the manipulations presented in Sections \ref{sec:linearsf} and \ref{sec:nonlinearsf}. After a straightforward but tedious calculation we find
\begin{align}
	m_{\rm eff} (\tau) a^\mu = {} & P^{\mu\nu} \bigg\{ \frac{ m^2 c_1^2 }{ m_{pl}^2 }  \Big[ f_\nu (z^\mu) + I_\nu ^{\rm tail} (z^\mu) \Big] \nonumber \\
	&- \frac{ m^3 c_1^2 c_2}{ m_{pl}^4 } \Big[ 2 f_\nu (z^\mu) I_{\rm tail} (z^\mu) + I_{\nu}^{\rm tail} (z^\mu) I_{\rm tail} (z^\mu) + \lim_{\epsilon \to 0^+} \int_{-\infty} ^{\tau - \epsilon} \!\!\! d\tau' \, \nabla_\nu D_{\rm ret} (z^\mu, z^{\mu'}) I_{\rm tail} (z^{\mu'}) \Big] \nonumber \\
	& + \frac{ m^4 c_1^2 c_2^2 }{ m_{pl}^6} \bigg[ f_\nu (z^\mu) I_{\rm tail}^2 (z^\mu) - \frac{1}{2\pi} f_\nu (z^\mu) \frac{ D I_{\rm tail} (z^\mu) }{ d\tau} - \frac{1}{4\pi} I_{\nu} ^{\rm tail} (z^\mu) \frac{ D I_{\rm tail} (z^\mu) }{ d\tau} \nonumber \\
	& {\hskip0.5in} + I_\nu ^{\rm tail} (z^\mu) \lim_{\epsilon \to 0^+} \int_{-\infty}^{\tau-\epsilon} \!\!\! d\tau' \, D_{\rm ret} (z^\mu, z^{\mu'}) I_{\rm tail} (z^{\mu'}) + I_{\rm tail} (z^\mu) \lim_{\epsilon \to 0 ^+ } \int_{-\infty}^{\tau- \epsilon} \!\!\! d\tau' \, \nabla_\nu D_{\rm ret} (z^\mu, z^{\mu'}) I_{\rm tail } (z^{\mu'}) \nonumber \\
	& {\hskip0.5in}+ 2 f_\nu (z^\mu) \lim_{\epsilon \to 0 ^+ } \int_{-\infty}^{\tau- \epsilon} \!\!\! d\tau' \,  D_{\rm ret} (z^\mu, z^{\mu'}) I_{\rm tail } (z^{\mu'}) - \frac{1}{4\pi} \lim_{\epsilon \to 0 ^+ } \int_{-\infty}^{\tau- \epsilon} \!\!\! d\tau' \, \nabla_\nu D_{\rm ret} (z^\mu, z^{\mu'}) \frac{D I_{\rm tail } (z^{\mu'}) }{ d\tau' } \nonumber \\
	& {\hskip0.5in} + \lim_{\epsilon \to 0^+} \int_{-\infty}^{\tau- \epsilon} \!\!\! d\tau' \, \nabla_\nu D_{\rm ret} (z^\mu, z^{\mu'}) \lim_{\epsilon' \to 0^+ } \int_{-\infty} ^{\tau' - \epsilon'} \!\!\! d\tau'' \, D_{\rm ret} (z^{\mu'}, z^{\mu''}) I_{\rm tail} (z^{\mu''}) \bigg]  \nonumber \\
	& + \frac{ m^4 c_1^3 c_3 }{ 2 m_{pl}^6} \bigg[ 2 f_\nu I_{\rm tail}^2 (z^\mu) + I_{\nu} ^{\rm tail} (z^\mu) I_{\rm tail}^2 (z^\mu) + \lim_{\epsilon \to 0^+} \int_{-\infty}^{\tau- \epsilon} \, \nabla_\nu D_{\rm ret} (z^\mu, z^{\mu'}) I_{\rm tail}^2 (z^{\mu'}) \bigg] + O(\varepsilon^4) \bigg\}
\label{renormalizedsf4}
\end{align}
where $I_{\rm tail}(z^\mu)$ and $I_{\nu}^{\rm tail}(z^\mu)$ are the tail integrals
\begin{align}
	I_{\rm tail} (z^\mu) & = \lim_{\epsilon \to 0^+} \int_{-\infty}^{\tau- \epsilon} \!\!\! d\tau' \, D_{\rm ret} (z^\mu, z^{\mu'}) \\
	I_\nu^{\rm tail} (z^\mu) & = \lim_{\epsilon \to 0^+} \int_{-\infty}^{\tau- \epsilon} \!\!\! d\tau' \, \nabla_\nu D_{\rm ret} (z^\mu, z^{\mu'}) ,
\end{align}
which are defined only on the worldline, and
\begin{align}
	f_\nu (z^\mu) = \frac{1}{4\pi} \bigg( \frac{1}{3} \frac{ D a_\nu }{ d\tau} + \frac{1}{6} R_{\nu \alpha} (z^\mu) u^\alpha \bigg)
\end{align}
is the local radiation reaction force in a non-vacuum background spacetime. The effective mass of the compact object is $m_{\rm eff} (\tau) = m \Gamma_{\rm Had} (z^\mu)$ where $\Gamma_{\rm Had} (z^\mu)$ is given by
\begin{align}
	\Gamma_{\rm Had} (z^\mu)  = {} & 1 - \frac{ m^2 c_1^2 }{ m_{pl}^2} I_{\rm tail} (z^\mu) + \frac{ m^3 c_1^2 c_2 }{ m_{pl}^4} \bigg[ \frac{1}{2} I_{\rm tail}^2 (z^\mu) + \lim_{\epsilon \to 0^+} \int_{-\infty} ^{\tau - \epsilon} \!\!\! d\tau' \, D_{\rm ret} (z^\mu, z^{\mu'}) I_{\rm tail} (z^{\mu'}) \bigg] \nonumber \\
	& - \frac{ m^4 c_1^2 c_2 }{ m_{pl}^6} \bigg[ I_{\rm tail} (z^\mu) \lim_{\epsilon \to 0^+} \int_{-\infty} ^{\tau - \epsilon} \!\!\! d\tau' \, D_{\rm ret} (z^\mu, z^{\mu'}) I_{\rm tail} (z^{\mu'}) - \frac{1}{4\pi} \lim_{\epsilon \to 0^+} \int_{-\infty} ^{\tau - \epsilon} \!\!\! d\tau' \, D_{\rm ret} (z^\mu, z^{\mu'}) \frac{D I_{\rm tail} (z^{\mu'}) }{ d\tau'} \nonumber \\
	& {\hskip0.65in} + \lim_{\epsilon \to 0^+} \int_{-\infty} ^{\tau - \epsilon} \!\!\! d\tau' \, D_{\rm ret} (z^\mu, z^{\mu'}) \lim_{\epsilon' \to 0^+} \int_{-\infty} ^{\tau' - \epsilon'} \!\!\! d\tau'' \, D_{\rm ret} (z^{\mu'}, z^{\mu''} ) I_{\rm tail} (z^{\mu''}) \bigg] \nonumber \\
	& - \frac{ m^4 c_1^3 c_3 }{ 2 m_{pl}^6} \bigg[ \frac{1}{3} I_{\rm tail}^3 (z^\mu)+ \lim_{\epsilon \to 0^+} \int_{-\infty}^{\tau- \epsilon } \!\!\! d\tau' \, D_{\rm ret} (z^\mu, z^{\mu'}) I_{\rm tail}^2 (z^{\mu'}) \bigg] + O(\varepsilon^4)  ~.
\label{effmass4}
\end{align}
Writing the self force in terms of the tail part of the propagator does not provide a condensed form as in (\ref{renormalizedsf3}) since the right side of (\ref{renormalizedsf4}) is not proportional to $\nabla_\nu \Gamma_{\rm Had} (z^\mu)$. This is because of the appearance of $f_\nu(z^\mu)$, which is not proportional to $\nabla_\nu$, and because the covariant derivative can not be pulled through the tail integrals in (\ref{renormalizedsf4}) without generating other terms that are not proportional to $\nabla_\nu$. However, the form in (\ref{renormalizedsf4}) and (\ref{effmass4}) clearly shows the role of tails-of-tails and tails-of-tails-of-tails in the self force and effective mass, respectively. 

The self force in (\ref{renormalizedsf4}) also depends explicitly on $D a_\nu/ d\tau$ so that an order reduction procedure should be implemented to exclude unphysical runaway and acausal solutions from appearing in the theory -- order reduction is a standard treatment of this issue in theories describing radiation reaction particle dynamics (see e.g., \cite{LandauLifshitz}).

\section{Conclusion}
\label{sec:conclusion}

In this paper we introduced a nonlinear scalar model for extreme mass ratio inspirals that is a natural analogue of the corresponding perturbative General Relativistic description. This model should be useful for studying the role of higher-order self force corrections for building high-accuracy waveforms for precise parameter estimation, for quantifying the effect of transient resonances on the phase evolution of waveforms, and for providing a sufficiently simple context to develop numerical methods for computations of higher-order self force corrections that may then be applied to the gravitational problem. This last program should be particularly useful for calibrating semi-analytic models \cite{BuonannoDamour:PRD59}, for building phenomenological-based hybrid waveforms \cite{Ajithetal:CQG24}, and for constructing gauge-invariant EMRI observables at higher orders in $\varepsilon$.

This model has the interesting feature that, despite being constructed to be as similar as possible to perturbed General Relativity, one can always perform a field redefinition using (\ref{psi1}) so that the initially nonlinear scalar field theory is transformed to a linear one (with suitable changes in the field-particle interactions). Such a transformation provides a very clean and simple derivation of the self-force corrections at higher orders since all contributions coming from the self-interaction of the field are subsequently removed. It is natural then to wonder if such a transformation can be made when describing gravitational EMRIs to remove some, if not all, of the nonlinear self-interactions of the metric perturbations. Indeed, it may be that a combination of a field redefinition and a gauge transformation (one that is $O(\varepsilon^2)$ so as to keep the linearized metric perturbations in the Lorenz gauge at leading order) successfully removes all nonlinear self-interaction terms at a given order. If this is the case then calculating higher order self force corrections will be much easier to derive analytically and solve numerically. We will discuss these issues further in a future paper.

In order to calculate the self force on the SCO we demonstrated how the usual action principle fails to retain the outgoing boundary conditions when integrating out the field -- a conservative dynamics for the worldline ensues because the full Lagrangian is time-reversal invariant. However, using the difference in the actions for two worldline and field histories (one evolving forward in time and the other backward) allows one to ``break'' the time-reversal symmetry naturally contained in the usual action principle and incorporate the time-asymmetric outgoing boundary conditions into the effective action. In this way, one can describe the open classical system dynamics of the SCO from an action principle and derive the self force (and waveforms -- see the next paper in this series) in a self-consistent manner.

Using this new action principle for open classical systems together with the effective field theory formalism we derived the self force on the SCO through {\it third} order in $\varepsilon$. Our main results are given in (\ref{renormalizedsf1}) and (\ref{renormalizedsf3}). We separated the singular parts of the self force from those that are regular on the worldline using the Detweiler-Whiting decomposition for the retarded Green's function. 
For completeness and comparison, we also presented the more complicated third-order self force expressions in terms of Hadamard's tail propagator in (\ref{renormalizedsf4}) and (\ref{effmass4}).
Despite the appearance of higher order poles and history-dependent divergent forces -- see (\ref{regselfforce1}) -- we found that only three {\it local} and {\it time-independent} counter terms are needed to absorb all of these divergences, which amounts to a renormalization of the SCO mass $m$ and the coupling constants $c_1$ and $c_2$ appearing in (\ref{nonlinear2}). 
In addition, we showed that this nonlinear scalar model only generates power-divergent integrals that automatically vanish in a regulariziation scheme that manifestly preserves the symmetries of the theory (e.g., dimensional regularization). No logarithmic divergences, which represent physical effects of screening by the scalar perturbations, can be generated in this model and thus none of the coupling constants appearing in (\ref{nonlinear2}) or, equivalently, (\ref{nonlinear1}) undergo a classical renormalization group running.

Our results in (\ref{renormalizedsf3}) suggest an ambiguity in defining the effective mass of the SCO since, in this model, the effective mass is proportional to the SCO's inertial mass ($m_{\rm eff}  = \Gamma_{\rm DW} m$) and one can associate the proportionality factor $\Gamma_{\rm DW} (z^\mu)$ with a contribution to the effective mass or with an additional contribution to the self force. Both interpretations yield the same acceleration on the SCO and the same waveform and physical observables.

In the next paper in this series we shall derive the expressions for the scalar perturbations emitted by the SCO through third order in $\varepsilon$. We shall also show that one can more easily derive the self force by calculating the regular part of the radiative field and then  evaluating it in the worldline equations of motion (see the discussion below (\ref{renormalizedsf3})). This suggests that current 3+1 self force codes \cite{Vegaetal:PRD80} may need only minimal modifications to compute higher-order self force corrections. In regularizing the third-order waveforms (which also contain divergences since the field is sourced by an effectively point-like object) we will show that exactly the same counter terms used in regularizing the third-order self force expressions -- see (\ref{counterterms10})-(\ref{counterterms12}) -- also regularize the scalar perturbations and the self force expressions derived from them. Therefore, the use of point particle treatments for the SCO in conjunction with our renormalization program
will be explicitly self-consistent.

\acknowledgments

We thank Tanja Hinderer, Michele Vallisneri, Ian Vega and Steven Detweiler for very helpful comments and discussions. 
This work was supported in part by an appointment to the NASA Postdoctoral Program at the Jet Propulsion Laboratory adminstered by Oak Ridge Associated Universities through a contract with NASA and in part by National Science Foundation grants PHY0801213 and PHY0908457.

\appendix

\section{Quasilocal expansions}
\label{app:expansions}

In this Appendix we provide the quasilocal expansions of various quantities used to regularize the expressions for the self force on the particle.

\begin{figure}
	\includegraphics[width=6cm]{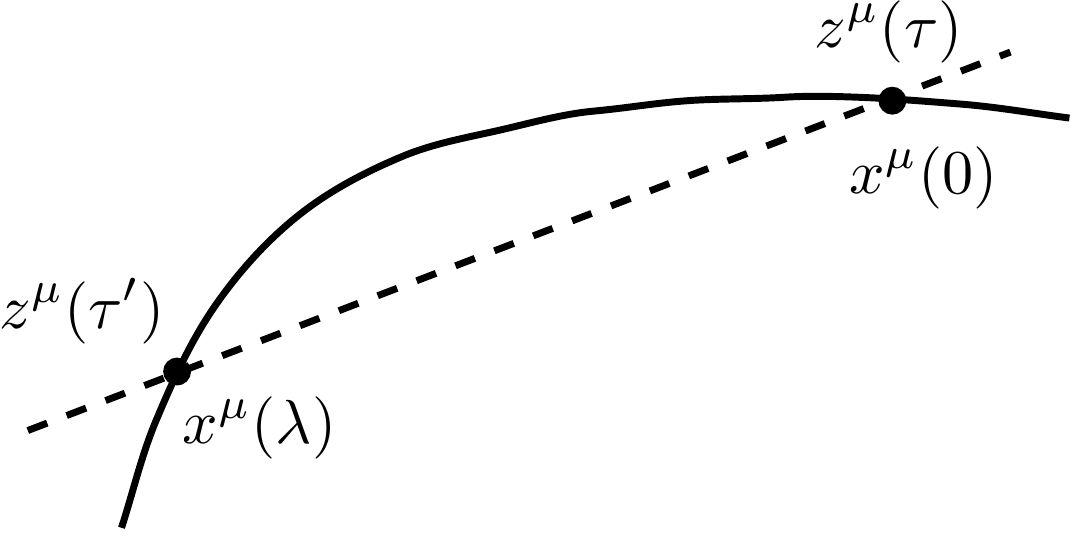} 
\caption{The relation between the worldline (solid line) and a geodesic (dashed line) intersecting the worldline at the coordinates $z^\mu(\tau)$ and $z^\mu (\tau')$ for constructing the quasilocal expansions of various quantities.}
\label{fig:quasilocal}
\end{figure}

For two points $z^\mu$ and $z^{\mu'}$ on the worldline the quasilocal expansions of the direct/singular part of the retarded propagator $\sim \Delta^{1/2} \delta (\sigma)$ follow from the expansions near coincidence (when $s \equiv \tau' - \tau$ is zero) of the van Vleck determinant $\Delta (x,x')$ and Synge's world function $\sigma (x,x')$. 
The quasilocal expansion of the covariant derivative of Synge's world function, $\sigma_\nu \equiv \sigma_{;\nu}$ can be derived by noting that a unique geodesic (with coordinates $x^\mu (\lambda)$ and affine parameter $\lambda$) intersects the accelerated worldline at the points $z^\mu (\tau)= x^\mu (0)$ and $z^\mu (\tau') = x^\mu (\lambda)$; see Figure \ref{fig:quasilocal}. For $\tau'$ near $\tau$ and $\lambda$ near $0$ it follows that $x^\mu (0) - x^{\mu} (\lambda) = z^\mu (\tau) - z^{\mu} (\tau')$ implies
\begin{align}
	- \lambda \, x'^\mu(0) - \frac{ \lambda^2 }{ 2 } x''^\mu (0) + \cdots = - s \dot{z}^\mu (\tau) - \frac{ s^2 }{ 2 } \ddot{z}^\mu (\tau) + \cdots
\end{align}
where a $'$ or $\dot{}$ denotes differentiation with respect to $\lambda$ or $\tau$, respectively. The geodesic equation can be used to eliminate terms with two or more parameter derivatives since 
\begin{align}
	x''^\mu(\lambda) = - \Gamma^\mu_{\alpha \beta} (x^\nu(\lambda)) x'^\alpha (\lambda) x'^\beta (\lambda) ~.
\end{align}
Finally, using the fact that $\sigma_\nu$ is proportional to the tangent on the geodesic at $\lambda=0$ so that $\sigma _\nu (x^\mu, x^{\mu'}) = - \lambda x'_\nu (0)$ \cite{Poisson:LRR} gives
\begin{align}
	\sigma_\nu (z^\mu, z^{\mu'}) & = -s u_\nu - \frac{ s^2 }{ 2!} a_\nu - \frac{ s^3}{ 3!} \frac{ D a_\nu }{ d\tau} + O(s^4)  ~.
\label{dsigma1} 
\end{align}
Sygne's world function is constructed from the identity $2\sigma = \sigma_\nu \sigma^\nu$,
\begin{align}
	\sigma (z^\mu, z^{\mu'}) & = - \frac{ s^2}{2} - \frac{ s^4 a^2 }{ 24} + O(s^5)  
\label{sigma1}
\end{align}
and the quasilocal expansion of $\delta (\sigma)$ then immediately follows,
\begin{align}
	\delta (\sigma (z^\mu, z^{\mu'}) ) & = \frac{ \delta (s) }{ | d\sigma / ds | } \\
		& = \frac{\delta (s) }{ | s | } \left( 1 + \frac{ 1}{6} s^2 a^2 + \frac{ 5 }{ 24} s^3 a^\alpha \frac{ Da_\alpha}{d\tau} + O(s^4) \right) ~.
\label{delta1} 
\end{align}
The covariant derivative of $\delta(\sigma)$ in the quasilocal expansion follows by noting that $\nabla_\nu = \sigma_\nu (ds/d\sigma) d/ds$. One can show that $ds/d\sigma = (d\sigma / ds)^{-1}$ and use the expressions in (\ref{dsigma1}) and (\ref{sigma1}) to find that
\begin{align}
	\nabla_\nu \delta (\sigma (z^\mu, z^{\mu'}) ) & = \sigma_\nu \frac{ ds }{ d\sigma} \frac{ d }{ ds } \delta (\sigma(z^\mu, z^{\mu'} ))   \\
		& = \left( u_\nu + \frac{ s}{2} a_\nu + \frac{ s^2}{ 6} P_{\nu \alpha} \frac{ D a^\alpha }{ d\tau} + O(s^3) \right) \frac{ d}{ds} \delta (\sigma(z^\mu, z^{\mu'} ))  ~.
	\label{ddelta1}
\end{align}
The quasilocal expansions of the van Vleck determinant and its covariant derivative follow from its expansion near coincidence \cite{Christensen:PRD14},
\begin{align}
	\Delta^{1/2} (x,x') = {} & 1 + \frac{ 1}{12} R_{\alpha \beta} \sigma^\alpha \sigma^\beta - \frac{1}{24} R_{\alpha \beta; \gamma} \sigma^\alpha \sigma^\beta \sigma^\gamma + \cdots 
\label{vvdet1}	\\
	\nabla_\nu \Delta^{1/2} (x,x') = {} & \frac{1}{6} R_{\nu \alpha} \sigma^\alpha - \frac{ 1}{24} ( 2 R_{\nu \alpha ; \beta} - R_{\alpha \beta; \nu} ) \sigma^\alpha \sigma^\beta + \cdots ~.
\label{vvdet2}
\end{align}
Inserting (\ref{dsigma1}) into these expressions gives their quasilocal expansions,
\begin{align}
	\Delta^{1/2} (z^\mu, z^{\mu'}) = {} & 1 + \frac{ s^2 }{ 12} R_{\alpha \beta} u^\alpha u^\beta + O(s^3) 
\label{vanvleck1}\\ 
	\nabla_\nu \Delta^{1/2} (z^\mu, z^{\mu'}) = {} & - \frac{ s}{ 6 } R_{\nu \alpha } u^\alpha + O(s^2)  ~.
\label{dvanvleck1}
\end{align}
These are all the expansions used in regularizing the self force corrections in this paper.

\section{Dimensional regularization}
\label{app:dimreg}

We have mentioned several times in the main body that the divergent integral $\Lambda$ vanishes in dimensional regularization. We will prove this here.

First, note that
\begin{align}
	\Lambda = \frac{1}{2} \int_{-\infty}^\infty ds \, \frac{ \delta (s) }{ | s | }
\end{align}
and from (\ref{delta1}) it follows that
\begin{align}
	\Lambda = \frac{1}{2} \int _{-\infty}^\infty ds \, \delta ( \sigma (z^\mu, z^{\mu'}) )   ~.
\label{Lambda2}
\end{align}
[In fact, it should be clear from (\ref{divintegral3}) that $\Lambda = 4\pi \int_{-\infty}^\infty ds \, D_S (z^\mu, z^{\mu'})$.] 
In Fermi normal coordinates, $\sigma(z^\mu, z^{\mu'}) = - s^2/2$ where $s = \tau' - \tau$. The integral only has support when $s \sim 0$ so that we can introduce a tangent space at the point $z^\mu(\tau)$ to define a Fourier representation for the integral in (\ref{Lambda2}). In particular, if $k^\mu$ is a ``momentum'' vector in this tangent space then
\begin{align}
	\Lambda & = \frac{1}{2} \lim_{m_\phi \to 0} \int_{-\infty}^\infty ds \int_{-\infty}^\infty \frac{ d^4k }{ (2\pi)^4 } \frac{ e^{ i k^0 s} }{ k^2 + m_\phi^2 }  \\
		& = \frac{1}{2} \lim_{m_\phi \to 0} \int_{-\infty}^\infty \frac{ d^3k }{ (2\pi)^3 } \frac{ 1}{ {\bf k}^2 + m_\phi^2 }
\label{Lambda3}
\end{align}
where we have included a small mass $m_\phi$ in the denominator to make the integral well-defined when $k^\mu \to 0$.

The divergence in evaluating the integral (\ref{Lambda3}) comes from those momenta with arbitrarily large magnitudes $k \to \infty$.
For a small enough spacetime dimension the ultraviolet behavior of $\Lambda$ can be improved so that the integral is finite. This is the motivation for dimensional regularization wherein the spacetime dimension is parameterized by a complex number $d$ so as to find a regime in the complex $d$-plane where the integral is finite. After evaluating the integral there, the result is analytically continued back to the physical four spacetime dimensions.

Here, the divergent integral (\ref{Lambda3}) in arbitrary dimensions is 
\begin{align}
	\Lambda = \frac{1}{2} \frac{\mu^{4-d} }{ (2\pi)^{d-1} }  \int_{S^{d-2}} \!\!\! d\Omega \lim_{m_\phi \to 0} \int_0^\infty dk \, \frac{ k^{d-2} }{ k^2 + m_\phi^2 } 
\end{align}
where $\mu$ is an arbitrary parameter that preserves the units of $\Lambda$.
Focusing on the last integral we see that
\begin{align}
	\lim_{m_\phi \to 0} \int_0^\infty dk \, \frac{ k^{d-2} }{ k^2 + m_\phi^2 } = - \lim_{m_\phi \to 0} \frac{ \pi \, m_\phi}{ 2}  ,
\end{align}
which obviously vanishes in the physical four dimensional spacetime. Therefore, $\Lambda = 0$ in dimensional regularization implying that one may replace $D_{\rm ret} (z^\mu, z^{\mu'})$ by $D_R (z^\mu, z^{\mu'})$ to obtain the correct regular self force, waveform, etc. as discussed in Section \ref{sec:renormalization} near (\ref{replacement1}).

\section{Feynman rules}
\label{app:feynmanrules}

The effective field theory approach \cite{GoldbergerRothstein:PRD73} utilizes Feynman diagram techniques that are useful for efficiently obtaining the effective action at a given order in perturbation theory. Specifically, writing down the connected and tree-level diagrams at a given order (see Section \ref{sec:pwrcounting}) is equivalent to solving the wave equation through that order and substituting that solution back into the action to get the contribution to the effective action at that order.

\begin{figure}
	\includegraphics[width=5cm]{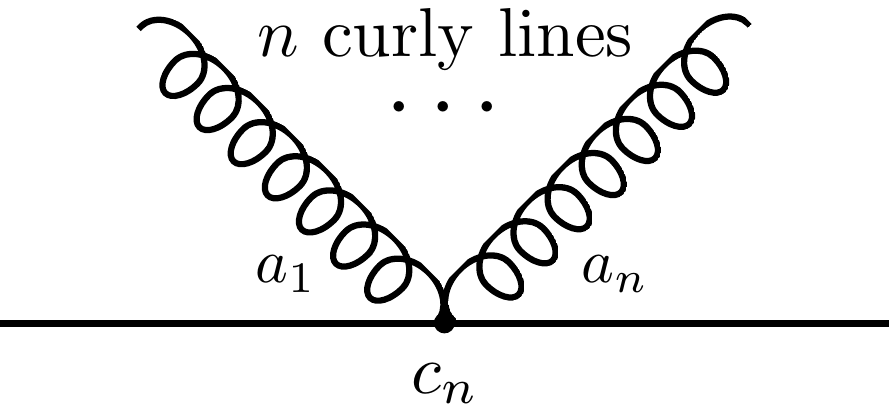}
\caption{The Feynman rule for the interaction of $n$ scalar fields on the worldline.}
\label{fig:wlvertex}
\end{figure}

There are rules, called Feynman rules, for translating Feynman diagrams into mathematical expressions. The Feynman rules needed for computing the self force, or any other quantity in the theory, are given by:
\begin{itemize}
	\item{For each worldline vertex represented by a dark circle and labeled by $a_1, \cdots, a_n$ and $c_n$ write down a factor of $T_{a_1 \cdots a_n} (x; z^\mu ]$. These are the worldline vertices.}
	\item{For each curly line write down a factor of $D^{ab} (x,x')$ connecting worldline vertices labeled by the indices $a$ and $b$ at spacetime points $x$ and $x'$ on the worldline.}
	\item{Sum over all labels $a_1 \cdots a_n$ in the chosen representation (e.g., the $\pm$ basis) and integrate over all spacetime points.}
	\item{Divide by the appropriate symmetry factor of the diagram.}
\end{itemize}
The symmetry factor for a given diagram is found by counting the number of ways one can permute the histories' indices (i.e., $a_1, a_2, \ldots$) while retaining the same set of propagators appearing in the diagram.
The vertex $T_{a_1 \cdots a_n}(x;z]$ comes from the interaction terms in the full action (\ref{nonlinear2}) involving a worldline integral over $n$ scalar fields,
\begin{align}
	- \frac{ m c_n }{ n! m_{pl}^n} \int d\tau \, \psi^n (z^\mu) = \frac{1}{n!} \int_x \psi^n (x) T (x; z^\mu]
\end{align}
where 
\begin{align}
	T(x; z^\mu] \equiv - \frac{ m c_n}{ m_{pl}^n } \int d\tau \, \frac{ \delta ^4 (x^\mu - z^\mu (\tau)) }{ g^{1/2} } = - \frac{ m c_n }{ m_{pl}^n} V(x; z^\mu]  ~,
\end{align}
which also defines $V(x; z^\mu]$. The indices $a_1, \cdots, a_n$ on $T_{a_1 \cdots a_n} (x; z]$ come from doubling the dynamical variables to ensure that the variational principle and the effective action are compatible with the outgoing boundary conditions on the scalar field as discussed in Section \ref{sec:inin}. Specifically, the contribution of this interaction term to that action is
\begin{align}
	\frac{1}{n!} \int_x \psi_1 ^n (x) T( x; z _1 ] - \frac{ 1}{n!} \int_x \psi_2 ^n (x) T(x; z_2] = {} & \frac{1}{n!} \int_x \psi^{A_1} (x) \cdots \psi^{A_n} (x) c_{A_1 \cdots A_n} {}^B T_B (x; z]
\end{align}
where $T_B(x; z] \equiv T(x; z_B]$, capital Roman letters take values in $\{1,2\}$ and 
\begin{align}
	c_{A_1 \cdots A_p} = \left\{ \begin{array}{cl}
							1 & A_1 = \cdots = A_p = 1 \\
							-1 & A_1 = \cdots = A_p = 2 \\
							0 & {\rm otherwise}
						\end{array} \right.
\end{align}	
In the $\pm$ basis, $c_{a_1 \cdots a_n}{}^b = \Lambda_{a_1} {}^{A_1} \cdots \Lambda_{a_n}{}^{A_n} \Lambda^b{}_B c_{A_1 \cdots A_n}{}^B$ where $\Lambda_a{}^A$ is given in Section \ref{sec:formaldevs}. Therefore,
\begin{align}
	T_{a_1 \cdots a_n} (x; z] = - \frac{ m c_n}{ m_{pl}^n} c_{a_1 \cdots a_n}{}^b V_b(x; z]
\end{align}
where $V_b (x; z] \equiv \Lambda_b {}^B V(x; z_B]$. This worldline vertex is represented by the diagram in Figure \ref{fig:wlvertex} and is all that is needed to construct the self force through third order in $\varepsilon$ in this nonlinear scalar model.

\bibliographystyle{physrev}
\bibliography{gw_bib}

\setlength{\parskip}{1em}

\end{document}